\documentclass[apj]{emulateapj}

\usepackage{natbib}
\usepackage{mathtools}
\usepackage{amsmath}
\usepackage{hyperref}
\usepackage{color}
\usepackage{multirow}

\defcitealias{merello_2013}{MM13b}
\defcitealias{merello_2013_alma}{MM13a}
\shorttitle{ALMA Observations of G331.512-0.103 II}
\shortauthors{Herv\'ias-Caimapo et al.}

\begin{document}

\title{ALMA Observations of the massive molecular outflow G331.512-0.103 II: physical properties, kinematics, and geometry modeling}

\author{Carlos Herv\'ias-Caimapo$^{1,2\star}$, Manuel Merello$^{3}$, Leonardo Bronfman$^1$, Lars \r{A}ke-Nyman$^4$, Guido Garay$^1$, Nadia Lo$^1$, Neal J. Evans II$^{5,6}$, Cristian L\'opez-Calder\'on$^4$, and Edgar Mendoza$^3$}

\affil{$^1$Departamento de Astronom\'ia, Universidad de Chile, Casilla 36-D, Santiago, Chile}
\affil{$^2$Jodrell Bank Centre for Astrophysics, School of Physics and Astronomy, University of Manchester, Oxford Road, Manchester M13 9PL, UK}
\affil{$^3$Universidade de S\~ao Paulo, IAG Rua do Mat\~ao, 1226, Cidade Universit\'aria, 05508-090, S\~ao Paulo, Brazil}
\affil{$^4$Joint ALMA Observatory (JAO), Alonso de C\'ordova 3107, Vitacura, Santiago, Chile}
\affil{$^5$Department of Astronomy, The University of Texas at Austin, 2515 Speedway, Stop C1400, Austin, TX 78712-1205, USA}
\affil{$^6$Korea Astronomy and Space Science Institute, 776 Daedeokdae-ro, Yuseong-gu, Daejeon, 34055, Republic of Korea}

\email{$^\star$E-mail: aconcagua.chc@gmail.com}

\begin{abstract}
We present observations and analysis of the massive molecular outflow G331.512-0.103, obtained with ALMA band 7, continuing the work from \citet{merello_2013_alma}. Several lines were identified in the observed bandwidth, consisting of two groups: lines with narrow profiles, tracing the emission from the core ambient medium; and lines with broad velocity wings, tracing the outflow and shocked gas emission. The physical and chemical conditions, such as density, temperature, and fractional abundances are calculated. The ambient medium, or core, has a mean density of $\sim 5\times 10^6$\,cm$^{-3}$ and a temperature of $\sim 70$\,K. The SiO and SO$_2$ emission trace the very dense and hot part of the shocked outflow, with values of $n_{\rm H_2}\sim10^9$\,cm$^{-3}$ and $T \sim 160-200$\,K. The interpretation of the molecular emission suggests an expanding cavity geometry powered by stellar winds from a new-born UCHII region, alongside a massive and high-velocity molecular outflow. This scenario, along with the estimated physical conditions, is modeled using the 3D geometry radiative transfer code \emph{MOLLIE} for the SiO(J$=8-7$) molecular line. The main features of the outflow and the expanding shell are reproduced by the model.
\end{abstract}

\section{Introduction} The formation of  massive stars is an important topic in stellar astrophysics that is open for debate. Despite the wealth of knowledge that has been obtained on young massive stellar sources, still there is no universal agreement on how massive stars are formed and evolve \citep{garay_1999,zinnecker_2007,kennicutt_2012}.

There is an accepted paradigm on how low-mass stars form and evolve \citep{shu_1987}. However, problems such as scarcity of young sources, large heliocentric distances (several kpc in some cases) and the lack of appropriate spatial resolution (that has been addressed with interferometers such as ALMA just in recent years) make the subject of formation of massive stars an open question. Two main theories compete on trying to explain this mechanism: the monolithic gravitational collapse/turbulent core accretion model \citep{mckee_2003}, which is basically the extension of the model of formation of low-mass stars to the massive ones, but with higher accretion rates and energetics; and the competitive accretion model \citep{bonnell_2001,bonnell_2004}, which states that low-mass seeds that will eventually form massive stars compete for the available gas in the gravitational potential of their native cluster.

Since massive stars have very short lifetimes, sources that are currently in the process of formation are very valuable. Molecular outflows are commonly detected toward protostellar objects, and observed physical properties such as outflow power, force and mass loss rate, seems to correlate over a large range of luminosities, suggesting that outflows found in massive star forming regions are a scaled-up version of those found in their low-mass star counterparts~\citep{tan14}. 
The injection of momentum and energy from outflows are important in star forming regions at small and large spatial scales, although the role of feedback at different stages of protostellar evolution is still not clear~\citep{fra14,bal16}. Thus, from an observational point of view, a principal objective is to identify and characterize examples of young, massive and powerful outflow-associated sources, as is the case for the work here.

The present work focuses on arcsecond-resolution ALMA observations of the massive molecular outflow G331.512-0.103 \citep{bronfman_2008}. This source is located in the tangent of the Norma spiral arm, at a heliocentric distance of $\sim 7.5$\,kpc, and corresponds to a bright MYSO object associated with the central region of the G331.5-0.1 giant molecular cloud \citep[][with a H$_2$ mass of $\sim 5 \times 10^6$\,M$_{\odot}$]{garcia_2014}, which shows evidence of ongoing massive star formation \citep[][which we will refer to as \citetalias{merello_2013}]{merello_2013}. There are several characteristics that make this object valuable and unique: the presence of very broad emission wings in CO, CS and SiO, which indicate the presence of a very powerful outflow; the compact emission \citep[not resolved at $8\farcs0$ resolution with APEX,][]{bronfman_2008}, which we interpret as the outflow lobes being closely aligned with the line of sight; an expanding bubble geometry, which indicates the presence of stellar winds possibly arising from the exciting star within the hyper compact HII region \citep[][which we will refer to as \citetalias{merello_2013_alma}]{merello_2013_alma}; and high energetics, which makes this source one of the most powerful massive outflows discovered. For example, in the compilation of outflows in O-type young stellar objects by \citet{lopez_sepulcre_2009}, only five objects have higher bolometric luminosity.

Part of the ALMA observations and the main results are summarized in \citetalias{merello_2013_alma}. The SiO(J$=8-7$) and CO(J$=3-2$) lines show broad velocity wings ($\pm 70$\,kms$^{-1}$). The SiO line shows ring-like emission, suggesting an expanding motion, probably a cavity being blown-out by the powerful stellar output from the central source. The H$^{13}$CO$^+$(J$=4-3$) line traces the systemic velocity structure, as well as the surrounding core with its narrow emission.

This work is the follow-up of \citetalias{merello_2013_alma}. Here we investigate for the first time 18 newly analyzed lines, along with the 4 already analyzed there, with the goal of characterizing the physical conditions, kinematics and morphology of the massive molecular outflow. In this new work, we analyze the SiO and SO$_2$ molecules, widely recognized as outflow tracers \citep{schilke_1997}; CH$_3$CCH, which thermalizes at densities of $\sim 10^4$\,cm$^{-3}$ and is a very good estimator of temperature \citep{fontani_2002,molinari_2016}; and SO isotopologues, which are tracing the hot-core chemistry \citep{charnley_1997,wakelam_2004}. The paper is organized as follows: Section \ref{sec:observations} describes the ALMA band 7 observations, giving the main parameters of the observed spectral lines. Section \ref{sec:results} presents the results of the observations, including integrated emission maps and position-velocity plots. Section \ref{sec:analysis} presents the analysis of physical conditions, geometry, and kinematics performed on the source. A 3D radiative transfer model of the SiO(J$=8-7$) line is also included. Our conclusions are summarized in Section~\ref{sec:conclusions}.
\section{Observations} \label{sec:observations} The observations were performed with the Atacama Large Millimeter/submillimeter Array (ALMA) during Cycle 0, as described in \citetalias{merello_2013_alma}. The primary beam was $17\farcs8$ and the synthesized beam was $1\farcs38 \times 0\farcs68$, with a position angle of $-37\arcdeg.6$. The interferometric observations miss the recovery of large spatial scales above $\sim 11-14\farcs0$.

The data were processed using the Common Astronomy Software Application \citep[CASA;][]{mcmullin2007}. The four spectral windows (SW) are centered at 345.8, 347.2, 357.3 and 358.6\,GHz, each extending over 1875\,MHz and consisting on 3840 channels. The generated maps considered a ``briggs'' weighting mode on the data (robust parameter = 0.5).

Figure~\ref{fig:spectral_windows}, in Appendix~\ref{appendix:complete_band}, shows the composite spectra of SW3-SW2 (top), and SW0-SW1 (bottom), integrated over a region of $12\arcsec \times 12\arcsec$ centered at $\alpha_{2000} = 16^h12^m10.09^s$, $\delta_{2000} = -51\arcdeg 28\arcmin 38\farcs4$. The identified lines are marked across both bands. The image shows several blended lines that required a careful channel-by-channel determination of the emission mask during the cleaning reduction process. The rich spectra exhibit sulphur-bearing, carbon chains and other complex molecules, characteristic of hot core line emission found toward other well-studied sources at similar frequency bands, such as Cepheus A East~\citep[e.g.,][]{brogan2007, brogan_2008}, Orion KL~\citep{schilke_1997_manuel}, and G5.89-0.39~\citep[e.g.,][]{hunter2008}.

\begin{figure*}
	\plottwo{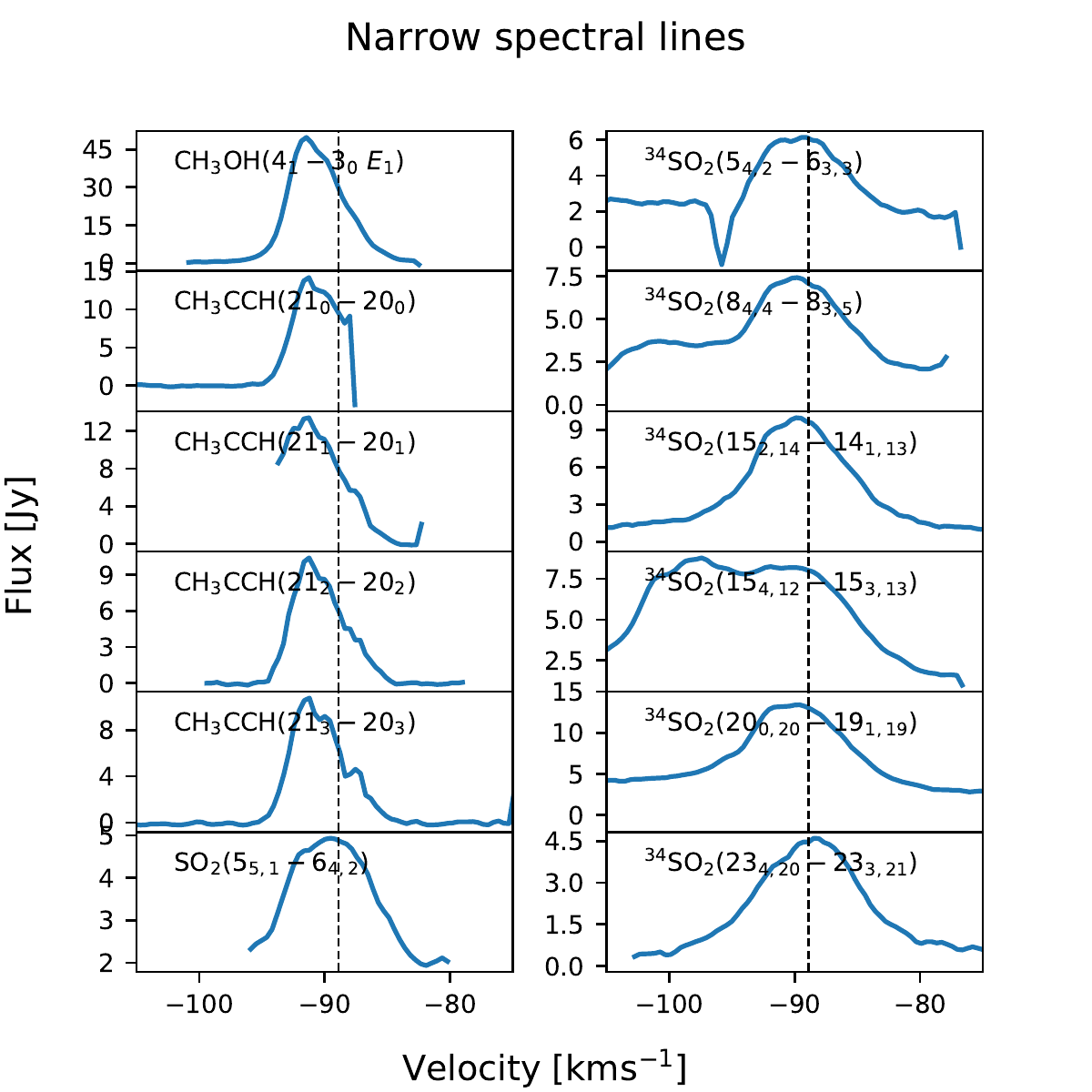}{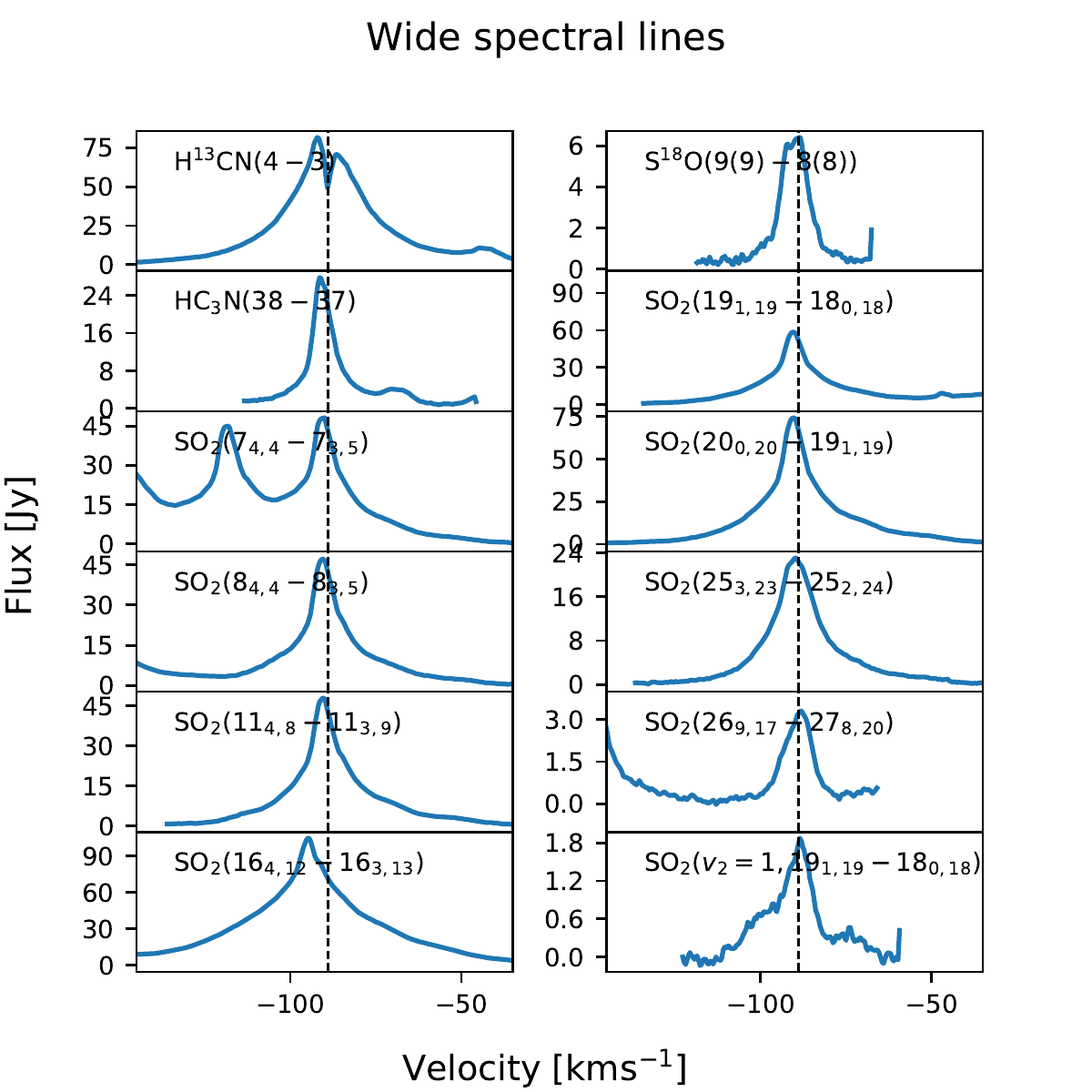}
	\caption{Integrated spectra of molecular lines discussed in this work and not presented in \citetalias{merello_2013_alma}. The left group corresponds to lines with narrow velocity emission, while the right group corresponds to lines with wide velocity emission. The dashed vertical lines shows the systemic velocity of $-88.9$\,kms$^{-1}$. \label{fig:integrated_spectra}}
\end{figure*}

For the present work, we focus on 18 (new) + 4(from \citetalias{merello_2013_alma}) lines, including, among others, the hot-core chemistry/shock tracer SO$_2$, a K-ladder of four CH$_3$CCH lines (commonly used as a source temperature test), the optically-thin line H$^{13}$CN probing densities up to $10^5$\,cm$^{-3}$, and HC$_3$N, which is found toward warm and dense regions such as hot cores, where it is shielded against destruction by photo-dissociation and C$^+$ ions \citep{rodriguez1998,prasad1980}. Figure~\ref{fig:integrated_spectra} shows the integrated spectrum of each of these lines as a function of velocity (the systemic velocity of the source is $-89$\,kms$^{-1}$). A couple of these lines appear blended or contaminated with emission at similar frequencies. We leave the analysis of the rest of the lines observed in the spectra, mostly associated with complex organic molecules such as dimethyl ether (CH$_3$OCH$_3$) and ethyl cyanide (C$_2$H$_5$CN), for upcoming studies.

The H$^{13}$CO$^+$($4-3$) line shows a secondary peak of emission at $\sim -50$\,kms$^{-1}$. \citetalias{merello_2013_alma} reports this as a ``molecular bullet'', or gas expelled at very high velocity by the powerful energetics of the outflow. Their presence has been reported previously in star formation region outflows \citep{tafalla_bachiller_2011}. However, once we analyzed the full band and the other lines contained in it, this might be no longer the case. A line with an excited vibrational level, HC$_3$N($J=38-37$,$v_7=1$) at 346.9491233\,GHz, matches with this secondary peak. Other star forming regions, observed at similar frequencies, show the presence of this line \citep{brogan_2008,nagy_2015}. We also observed, in our band-7 data, the HC$_3$N($J=38-37$) line at 345.60901\,GHz. Comparison between them in velocity and morphology might indicate that both belong to emission of the same molecule, so the presence of a molecular bullet can be discarded, instead appearing to be another line that was not recognized in the first analysis of the ALMA data.

\section{Results} \label{sec:results}

\subsection{Moment 0 maps} \label{sec:moment_0_maps} 
\begin{figure*}
	\plotone{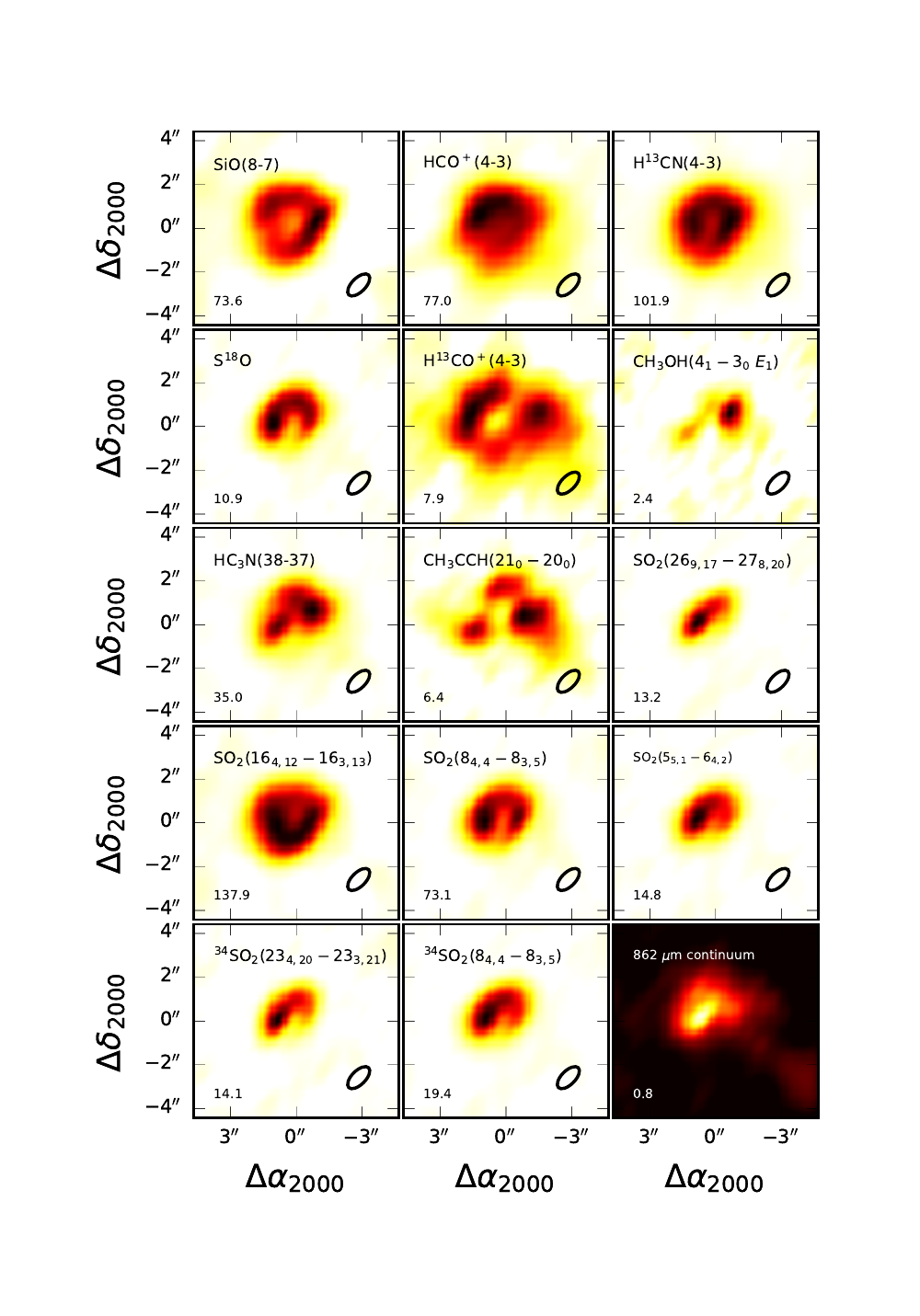}
	\caption{Integrated emission maps at systemic velocity of some representative lines in this work. The peak integrated intensity is shown in each panel in units of Jybeam$^{-1}$. The bottom right panel shows the 862\,$\mu$m continuum emission map. The ($0\farcs0$,$0\farcs0$) offset corresponds to $16^h12^m09^s.99$ $-51^{\circ} 28' 37\farcs75$. \label{fig:ambient_maps}}
\end{figure*}

Figure~\ref{fig:ambient_maps} shows imaging of the 0th moment maps of emission, integrated in the velocity range $-95.6$ to $-79.9$\,km\,s$^{-1}$ (systemic velocity range). 

In the integrated spectra, from Figure~\ref{fig:integrated_spectra}, we can identify two groups of lines. The first group (labeled \emph{wide} lines) are the ones with broad velocity wings. They are the SiO, S$^{18}$O, HCO$^+$, H$^{13}$CN, HC$_3$N, and most of the SO$_2$ lines. At the systemic velocity, most of these lines, especially the SiO emission, trace an emission with a ring-like shape (see Figure~\ref{fig:ambient_maps}).

The second group (labeled \emph{narrow} lines) includes lines that have narrow emission at systemic velocity. They are the CH$_3$CCH, CH$_3$OH, and H$^{13}$CO$^+$ lines \footnote{H$^{13}$CO$^+$ also shows evidence of high-velocity wings, but with a signal-to-noise ratio $\lesssim 2$}.

Spatially, the emission from all lines comes roughly from the same central region, shown in Figure~\ref{fig:ambient_maps}, with a diameter of $\sim 5\farcs0$. Most of the lines trace the ring-like emission that is clearly seen at the systemic velocity.


The 862\,$\mu$m continuum emission is shown in Figure~\ref{fig:ambient_maps} as the inverted color map in the bottom right corner. It is composed of a single peak that almost coincides with the center of the SiO emission ring, and weaker emission that follows the edge of the ring feature.



\begin{turnpage}
\begin{figure*}
	\plottwo{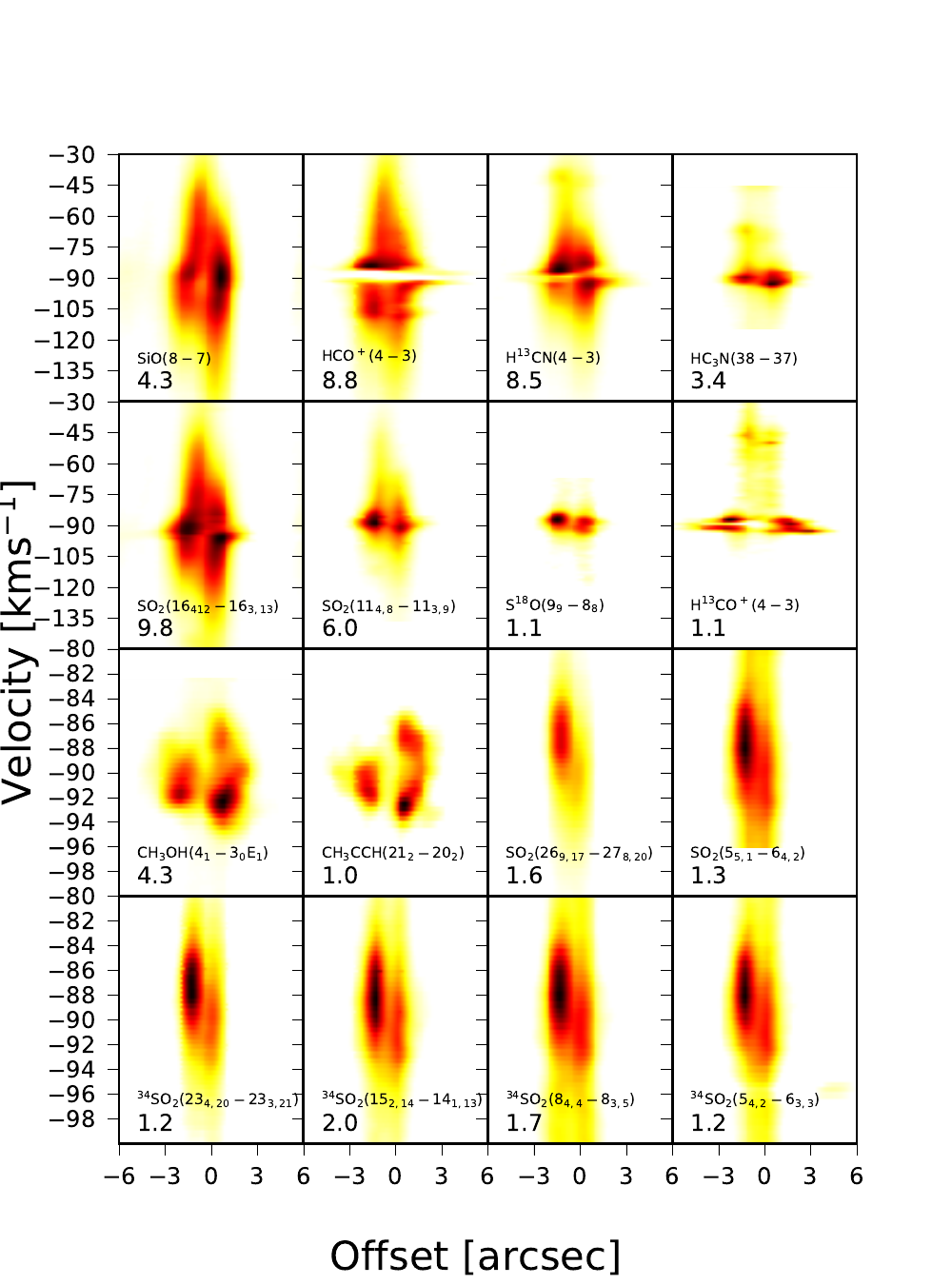}{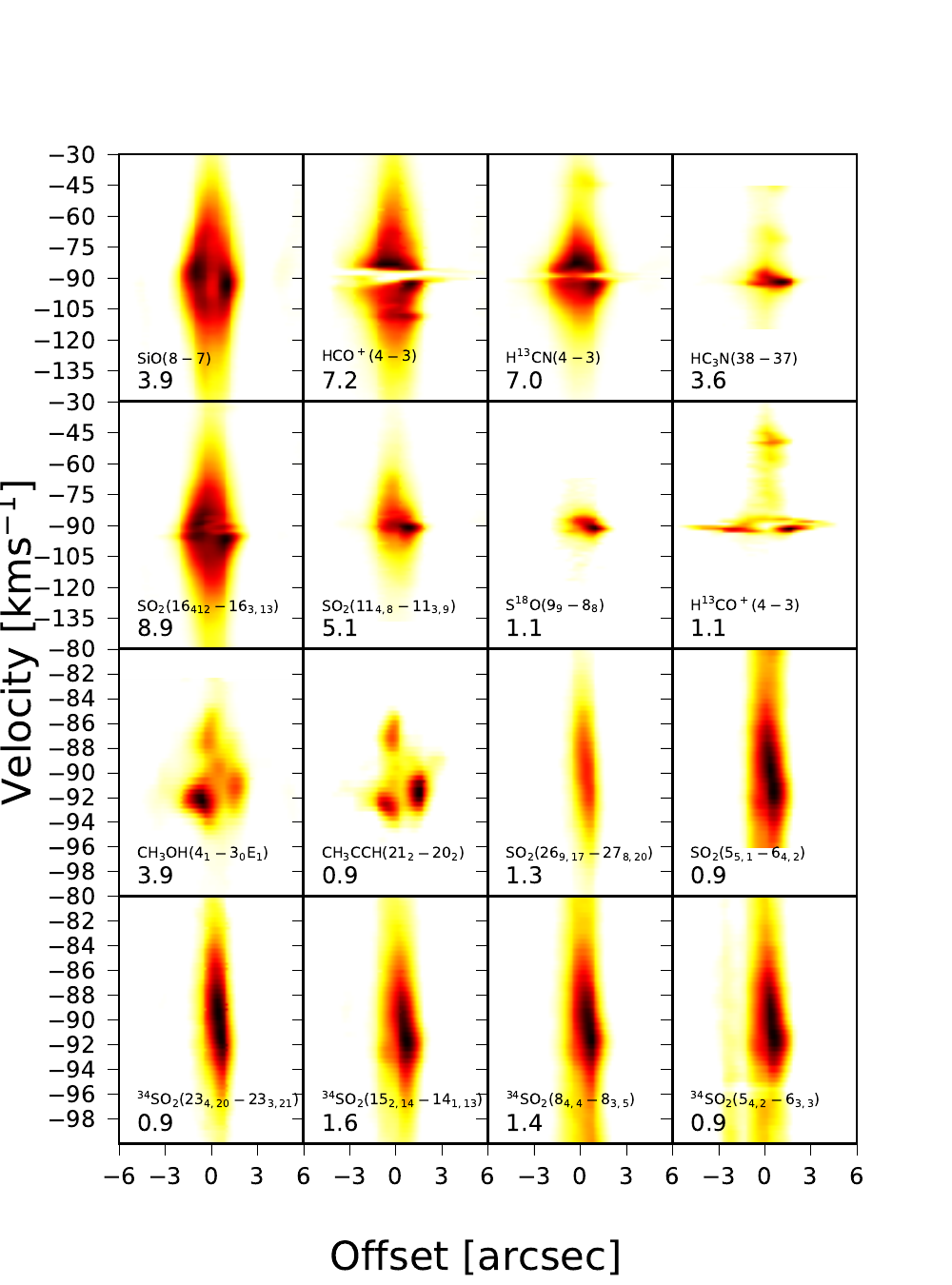}
	\caption{Left: PV plots for several observed lines along the outflow axis. The name of the line and the peak emission is shown. Right: PV plots for several observed lines perpendicular to the outflow axis. The name of the line and the peak intensity is shown, in units of Jybeam$^{-1}$. \label{fig:pv}}
\end{figure*}
\end{turnpage}

\subsection{Position-velocity (PV) plots} To study the gas kinematics traced by the line emission, we made PV plots of most of the observed lines along two axes: parallel and perpendicular to the outflow axis with a position angle of $102^{\circ}.5$, as defined in \citetalias{merello_2013_alma}. A ``slit'' of 11 pixels ($\sim 1\farcs5$) is used. Both kinds of PV plots are shown in Figure~\ref{fig:pv}.



\section{Analysis} \label{sec:analysis}

\begin{figure}
	\plotone{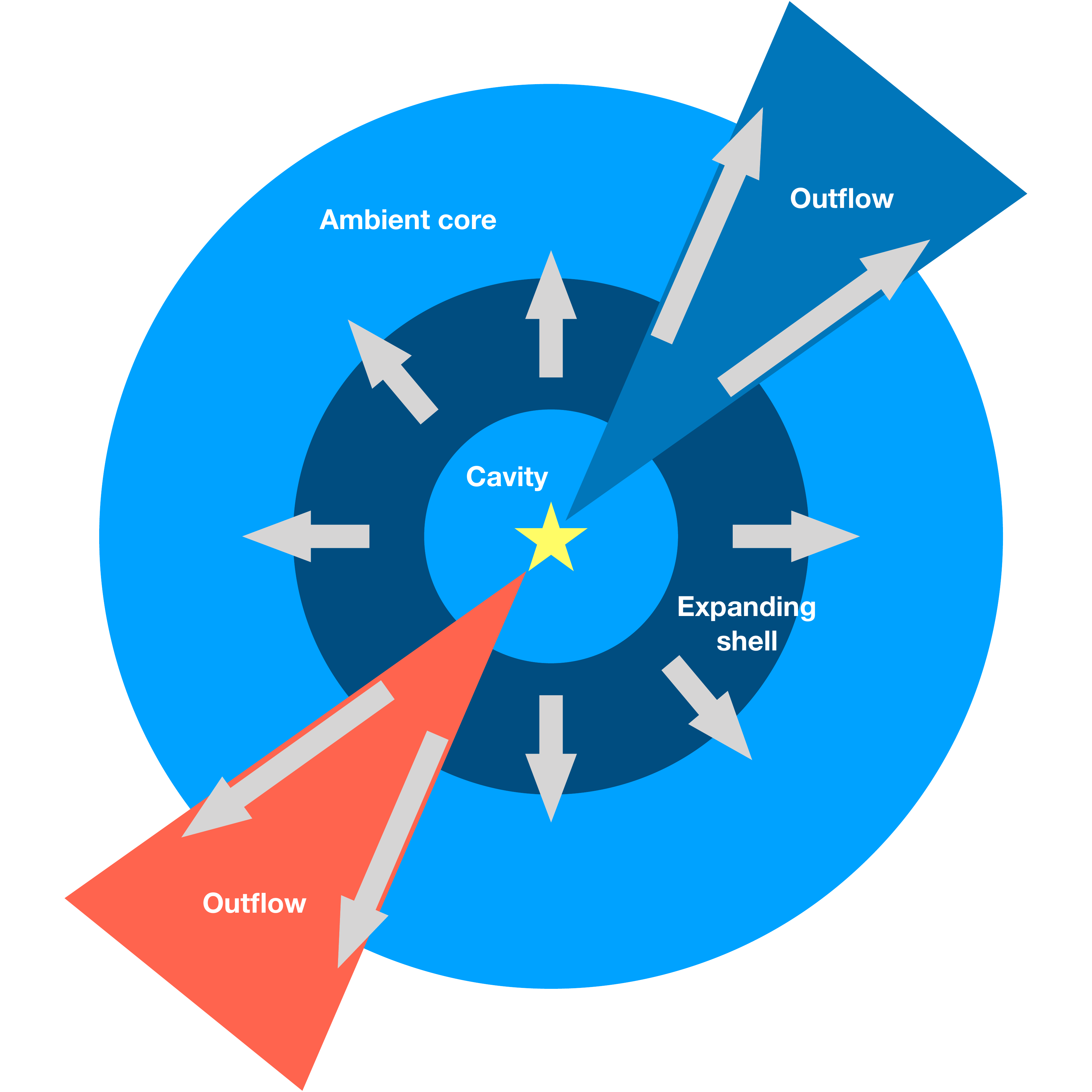}
	\caption{Working model of the G331 .512-0.103 massive core and outflow. The main morphological features are shown. \label{fig:sketch}}
\end{figure}                   

We interpret the wide group of lines as tracers of shocked high-velocity hot gas. The narrow group of lines trace the ambient core cold gas emission at the systemic velocity. As reported in \citetalias{merello_2013_alma}, we interpret the kinematics seen in the PV plots of SiO as an expanding shell. 

We present our interpretation of the source, shown in Figure~\ref{fig:sketch}. The bipolar outflow is outlined by emission from shock tracers, such as SiO and SO$_2$. We consider the presence of an expanding shell, which we interpret as shocked gas material being blown out by the stellar wind from the ultra compact HII region at the center. The ring-like emission we see in the maps of the lines would correspond to this expanding shell. The narrow group of lines traces emission at systemic velocities (e.g. CH$_3$CCH) and the ambient dense and warm gas core.



\subsection{Physical conditions} \label{sec:physical_conditions}

\begin{deluxetable*}{l c c c c}
\tablecaption{Physical conditions estimates on the averaged spectra. \label{table:physical_conditions}}

\tablehead{ 
		\colhead{Location}  & \multicolumn{2}{c}{Using SiO} 	& \multicolumn{2}{c}{Using SO$_2$}	\\
		\colhead{} 			& \colhead{Column density} 	& \colhead{Rot. temperature} & \colhead{Column density} & \colhead{Rot. temperature} \\
		\colhead{}          & \colhead{blue/red wing} 		& \colhead{blue/red wing} & \colhead{blue/red wing} & \colhead{blue/red wing} \\
		\colhead{} 			& \colhead{[$10^{14}$ cm$^{-2}$]} 			& \colhead{[K]} 	& \colhead{[$10^{16}$ cm$^{-2}$]} & \colhead{[K]} 
 }
			
\startdata

50\% emission peak 				&$9.1\pm0.1$/$8.3\pm0.1$    &	\multirow{4}{*}{$123\pm15$/$138\pm29^{\dagger}$}	&$6\pm1$/$3.8\pm0.5$    & $163 \pm 22$/$169 \pm 18$	\\
cavity							&$9.5\pm0.1$/$14.4\pm0.1$	&		&$8\pm2$/$8\pm1$		& $168 \pm 36$/$193 \pm 28$ \\
blue peak  						&$14.5\pm0.1$/$10.0\pm0.1$	&		&$9\pm1$/$6\pm1$		& $146 \pm 12$/$147 \pm 17$ \\
red peak 						&$7.8\pm0.1$/$15.1\pm0.1$	& 		&$8\pm3$/$8\pm2$		& $189 \pm 47$/$202 \pm 28$ \\	

\enddata
\tablecomments{$\dagger$ This values are estimated in \cite{merello_2013} with APEX data, which covers the source completely. Therefore, this is a rough estimate.}

\end{deluxetable*}


\subsubsection{Estimation of column density and temperature} \label{sec:rot_dia} 

\emph{SiO analysis - } Assuming a common $T_{\rm rot}$ for the SiO transitions and that they are optically thin, we calculate the column densities of the outflow wings. See, for example, the analysis in the similar outflow source W51 North \citep{zapata_2009}. In this ALMA data set, we only have observed one line of SiO at high resolution. Therefore, for the excitation temperature value, we consider the estimate from \citetalias{merello_2013}, using only two transitions of SiO, obtained with APEX, for each outflow wing: $123\pm15$\,K (blue wing) and $138\pm29$\,K (red wing).

The estimated averaged column densities are listed in Table~\ref{table:physical_conditions}. The significant values are the ones calculated in the blue and red peak, $1.45\times10^{15}$\,cm$^{-2}$ and $1.51\times10^{15}$\,cm$^{-2}$, respectively, which are taken as the column densities of the outflow wings. Estimating the size of the emission from the 50\% peak contour level in the 0th moment maps in the blue and red peaks, and using an abundance of $X_{\rm SiO}=1.3 \times 10^{-8}$ (see Section~\ref{sec:abundances}), we estimated the masses to be $\sim 12 M_{\odot}$, for each outflow wing. This value is about half of the mass calculated for the wings in \citet{bronfman_2008}, with CO.

\emph{SO$_2$ analysis - } One of the principal molecules observed in the ALMA data set corresponds to SO$_2$, since several transitions fall in the observed bandwidth. In this case, assuming optically thin emission, we use the rotational diagram technique \citep{linke_1979,goldsmith_1999} to obtain its column density and rotational temperature. 


The partition function for SO$_2$ is \citep{claude_2000} 
\begin{equation} 
Q(T_{rot}) = \frac{5.34 \times 10^{16}}{2} \sqrt{\frac{(T_{\rm rot}/{\rm K})^3}{ABC/{\rm MHz}^3}}  \text{,}
\end{equation}
where $A$=60778.5511\,MHz, $B$=10317.96567\,MHz and $C$=8799.80750\,MHz.

Since some SO$_2$ lines are blended, a compromise must be reached between the range of the wings to be used and the number of lines. In the ideal case, one wants to include the complete width of the wings and use as many lines as available. Using a narrow range of velocities is not advisable since the column density is underestimated. Two SO$_2$ lines (J=$5_{5,1}-6_{4,2}$ and J$=7_{4,4}-7_{3,5}$) were discarded since they are blended. Another line (J$=16_{4,12}-16_{3,13}$) has a clear excess over the other lines, and may be contaminated by an unknown line. In total, 7 lines were used in the blue and red wings.

\begin{figure}
	\plotone{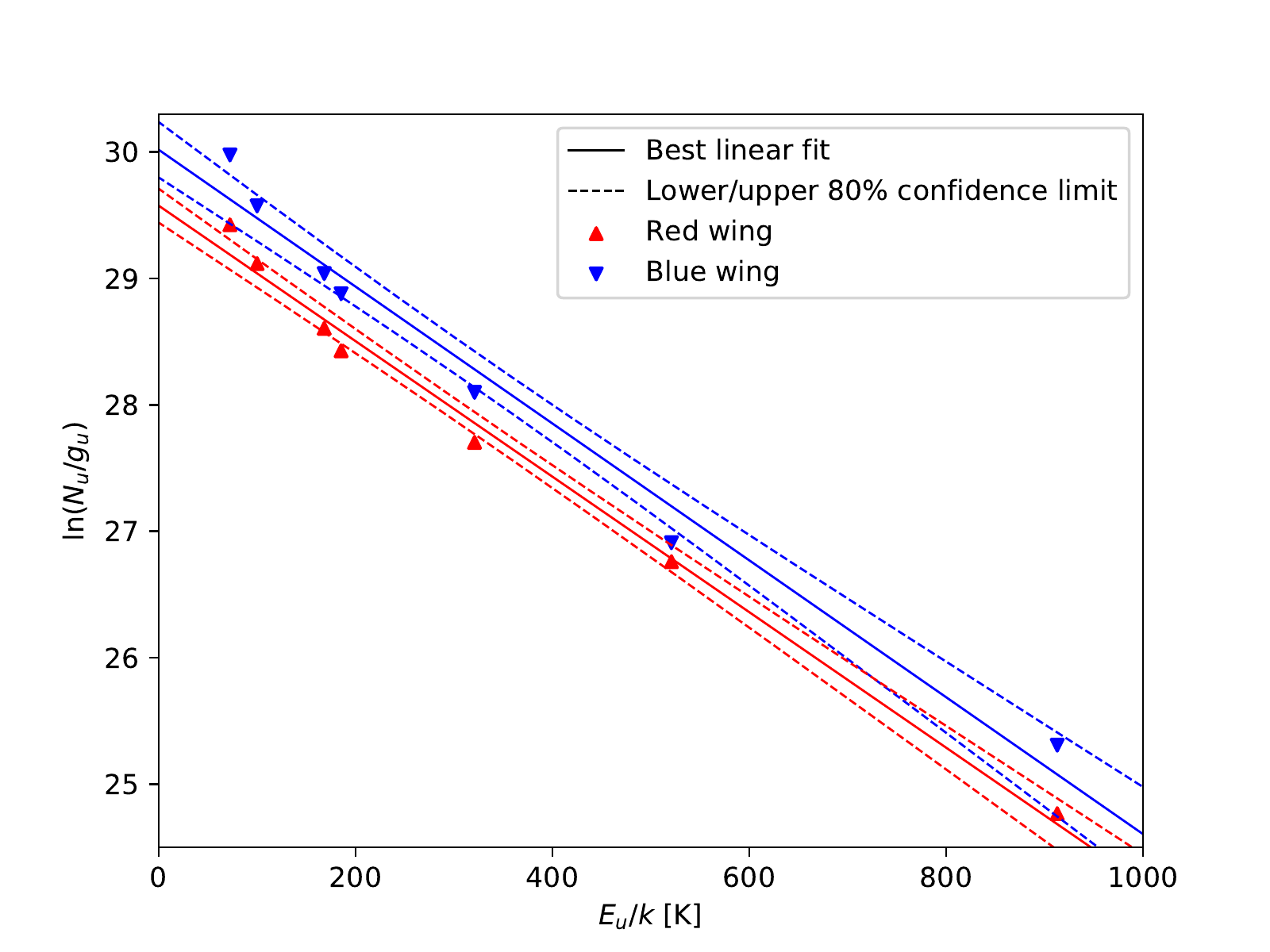}
	\caption{Example rotational diagram of the SO$_2$ lines, corresponding to the averaged spectra inside the 50\% of the peak emission contour. The blue and red points correspond to the spectrum integrated in the blue and red wing. The solid straight line corresponds to the linear regression fit. The dashed lines are the 80\% confidence intervals.\label{fig:rd_so2}}
\end{figure}

We consider the spatially averaged spectrum inside the contour of the 50\% of the peak emission, as well as inside a circle with $1\farcs0$ of diameter (corresponding to one beam) centered in the blue wing peak, the red wing peak, and the cavity center. These circles average roughly 40 pixels from the original data cube. With the averaged spectra in these four locations, we calculate the rotational diagrams and fit a straight line with a linear regression, from which we estimate column densities and rotational temperatures, listed in Table~\ref{table:physical_conditions}. The 1$\sigma$ errors are calculated from the covariance matrix from the linear fit. The blue and red peaks show the highest column density, of $9 \times 10^{16}$ and $8 \times 10^{16}$\,cm$^{-2}$, respectively. This is consistent with the same tendency observed in SiO. The average rotational temperature in the SO$_2$ emitting region is $\sim 170$\,K. The red peak is hotter than the blue peak, 202 and 146\,K, respectively. As an example, in Figure~\ref{fig:rd_so2}, we show the SO$_2$ rotational diagram for the flux averaged inside the 50\% peak emission contour. For the linear fit, we consider a conservative error of 20\% for the $\ln(E_u/g_u)$ values. We show the 80\% confidence intervals.

\begin{figure}
	\plotone{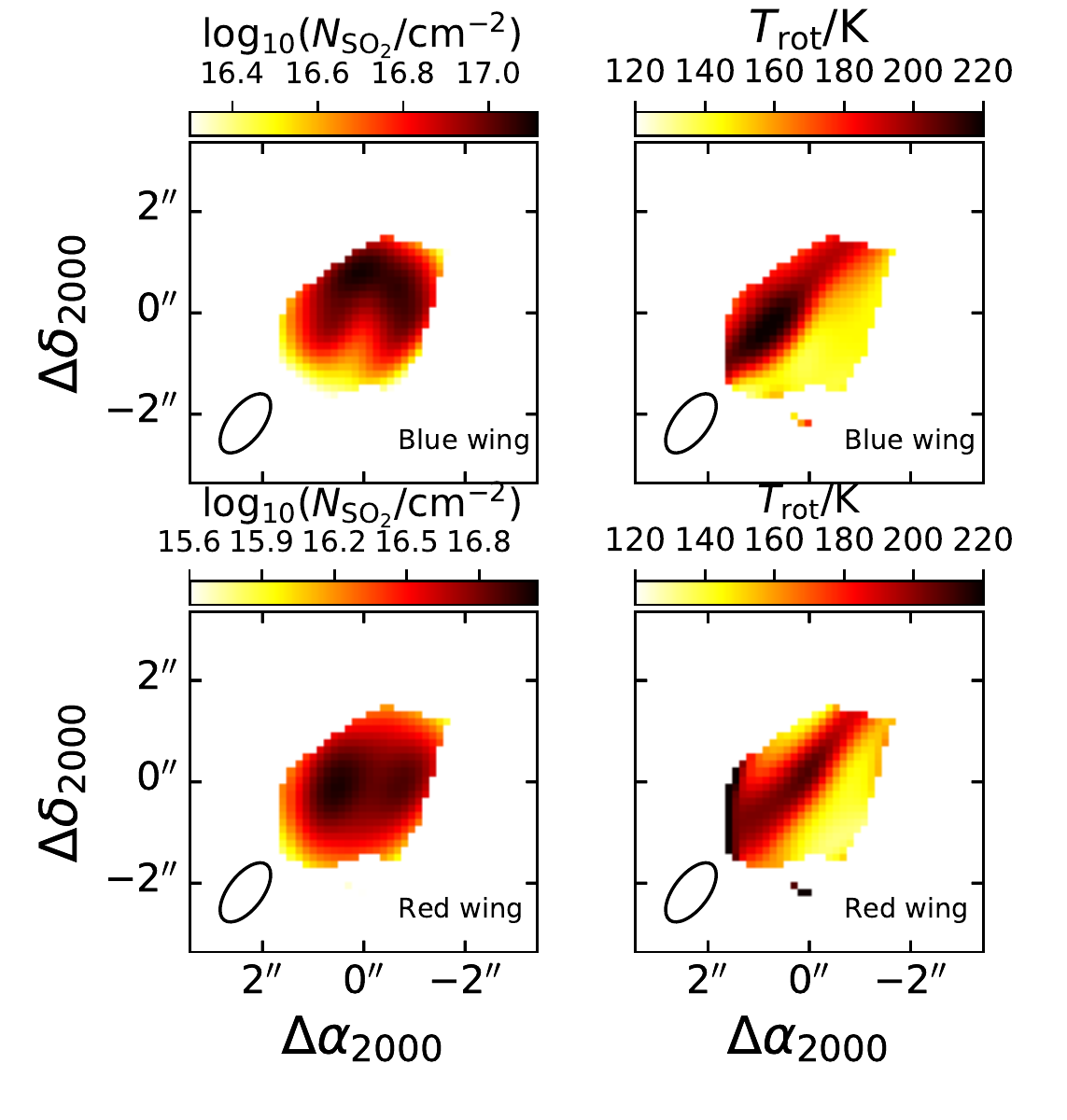}
	\caption{Column density (left column) and rotational temperature (right column) maps made with SO$_2$ rotational diagrams. The blue and red wings velocity ranges are $-120$ to $-95.6$ and $-79.9$ to $-67$\,km\,s$^{-1}$.\label{fig:cd_rt_so2}}
\end{figure}

The estimated SO$_2$ column density and rotational temperature maps for each outflow wing are shown in Figure~\ref{fig:cd_rt_so2}. The ring-like emission is traced in SO$_2$ by the blue and red wing emission. The rotational temperature maps show an elongated peak at $>200$\,K on both wings. The bright peak of emission, seen in most of the SO$_2$ lines, corresponds to the peak in temperature in the blue wing, and also seen in $N_{\rm SO_2}$ in the red wing. This means that SO$_2$ is likely tracing the most dense, hot, and shocked part of the proto-stellar core.

\begin{deluxetable*}{lccccc}[t]
		\tablecaption{Range of abundances for the SiO and SO$_2$ molecules. Fractional abundances in other massive outflow sources. For reference, the spatial resolution of our ALMA data is $\sim 10$\,kAU. \label{table:abundances}}
		\tablehead{ \colhead{} & \colhead{G331, $T_d$=70 K} 	& \colhead{G331, $T_d$=400 K} & \colhead{G5.89-0.39} & \colhead{Orion-KL} & \colhead{G34.26+0.15} }		
		\startdata
			SiO 			& $1.7(-9)$ 	& $1.3(-8)$		&	$3(-10)^a$ & $2.2(-8)^b$ & $\sim 1(-8/9)^d$\\
			SO$_2$			& $1.5(-7)$		& $7.5(-7)$ 	&	$2(-8)^a$ & $4.8(-8)^c$ & $1.4-3.7(-8)^e$ \\
		\enddata			
		\tablecomments{ 
        	$^a$ \citet{klaassen_2006}, JCMT, spatial resolution $\sim 30$\,kAU \\
  			$^b$ \citet{tercero_2011}, IRAM 30m, s. r $\sim$ 4-12\,kAU \\
  			$^c$ \citet{esplugues_2013}, IRAM 30m, s. r. $\sim$ 4-12\,kAU \\
  			$^d$ \citet{hatchell_2001}, VLA, s. r. $\sim$ 10 - 84\,kAU \\
  			$^e$ \citet{mookerjea_2007}, BIMA, s. r. $\sim$ 4\,kAU}
\end{deluxetable*}

\subsubsection{Fractional abundances} \label{sec:abundances} One physical parameter that is hard to measure, yet needed for modeling is the fractional abundance of each molecule. The fractional abundance $X_M$, where $M$ stands as a particular molecule, is measured as $X_M = \frac{N_M}{N_{H_2}}$ where $N_M$ is the column density of the correspondent molecule.

In order to calculate the fractional abundances, some evaluation of the mass is needed. The virial mass can be used to estimate it \citep[e.g. ][]{shirley_2003}, or the mass estimated from the thermal continuum emission, the \emph{dust mass}, can also be used \citep[e.g. ][]{fontani_2002}. The latter method is used in this work, and the mass is calculated from the 862\,$\mu$m continuum emission.

The equation to calculate the gas mass from dust thermal continuum, when the emission is optically thin, is
\begin{equation}
	M_{\rm d} = \frac{S_{\nu}D^2R}{\kappa_{\nu}B_{\nu}(T_{\rm d})}\text{,}
\end{equation}
where $S_{\nu}$ is the observed flux, $\kappa_{\nu}$ is the dust mass coefficient, $D$ is the distance to the source, $B_{\nu}$ is the Planck function, $T_{\rm d}$ is the dust temperature and $R$ is the gas to dust ratio, which is uncertain, but usually set to 100 in the literature. The following values will be used: $\kappa_{\nu}=1.89$\,cm$^2$g$^{-1}$ \citep[interpolated from model OH5 in][]{ossenkopf_1994} and $D=7.5$\,kpc. On the other hand, the gas mass ($M_{\rm cd}$) can be estimated from

\begin{equation}
	M_{\rm cd} = \frac{A N_{M} \mu m_{\rm H}}{X_M} \text{,}
\end{equation}

\noindent where $\mu m_{\rm H}$ is the mass of the hydrogen atom times the mean molecular weight (that accounts for a fraction of helium, equal to 2.29), $A$ is the area of the emitting region (which must be consistent with the area used to calculate the dust flux $S_{\nu}$) and $N_M$ corresponds to the measured column density of the molecule $M$. Equating $M_{\rm cd}$ to $M_{\rm d}$, and using the fact that $\Omega = A/D^2$ corresponds to the solid angle that the source subtends, we have

\begin{equation} X_M = \frac{\Omega N_M \mu m_{\rm H} \kappa_{\nu} B_{\nu}(T_{\rm d})}{S_{\nu} R} \text{.} \end{equation}

One quantity that is highly uncertain and sensitive for the dust mass estimation is the dust temperature. \citetalias{merello_2013} estimates the dust temperature to be 35\,K, based on the Spectral Energy Distribution (SED) of the source. This value is for the entire clump though, so the dust temperature at the ALMA resolution is probably higher. A rotational diagram analysis of the CH$_3$CCH lines shows that the ambient core temperature is $\sim 70$\,K. \citetalias{merello_2013_alma} uses an equilibrium dust temperature of 400\,K within the central arcsec. We use these two limits to calculate a range of possible abundances, listed in Table~\ref{table:abundances}.

We compare our fractional abundances results with similar sources studied in the literature. As stated in \citetalias{merello_2013_alma}, one of the most similar sources to G331.512-0.103 is G5.89-0.39, an UCHII region which exhibits a powerful and compact bipolar molecular outflow. \citet{klaassen_2006} mapped this outflow with the James Clerk Maxwell Telescope (JCMT) at $\sim 15\farcs0$. The fractional abundances for SiO and SO$_2$ are listed in Table~\ref{table:abundances}, column 4. The Orion-KL region \citep{genzel_1989} has shown chemical richness. \citet{tercero_2010} and references therein, reported a spectral line survey in the 90-300\,GHz range. We list their fractional abundances, for comparison, in Table~\ref{table:abundances}, column 5. Similarly, the UCHII region G34.26+0.15 \citep{garay_1986,wood_1989} is also a well studied massive star forming source. For comparison, the fractional abundances are listed in Table~\ref{table:abundances}, column 6.

\subsubsection{Density and column density estimation with \emph{RADEX}} \label{sec:radex} 

\begin{deluxetable*}{l l l l l}[b]
\tablecaption{Points probed in SO$_2$ lines to test the $\chi^2$ minimization with RADEX and the estimated properties. \label{table:all_points}}
\tabletypesize{\tiny}
\tablehead{ \colhead{Location} & \colhead{$\alpha$} & \colhead{$\delta$} & \colhead{$\log_{10}(n)$} & \colhead{$\log_{10}(N(\rm{SO}_2))^{\dagger}$} \\             \colhead{} & \colhead{J2000} & \colhead{J2000}	& \colhead{[cm$^{-3}$]}	& \colhead{[cm$^{-2}$]}	\\}
\startdata
SiO blue peak	  	    & 16:12:09.91		& -51:28:37.4	& $9.4$		& $16.5$	\\
SiO red peak			& 16:12:10.08		& -51:28:37.6	& $9.7$		& $16.6$ 	\\
SiO cavity center		& 16:12:09.99		& -51:28:37.5	& $9.3$		& $16.6$	\\
\enddata
\end{deluxetable*}

The radiative transfer code \emph{RADEX} \citep{vandertak_2007} is a non-LTE radiative code that uses radiative/collision rates information and basic geometries to solve the statistical equilibrium and radiative transfer equations, estimating the intensities of several transitions of the most typical molecules observed in the sub-mm range.

Since the statistical equilibrium/population levels and the radiative transfer/radiation field are coupled, \emph{RADEX} uses the Large Velocity Approximation (LVG) approximation by \citet{sobolev_1960}, where the mean intensity $\bar{J}$ is expressed as a function of the source function $S$ and a photon escape probability $\beta$. All the knowledge of the geometry/optical depth goes into this probability $\beta$. An estimate is given by 
\begin{equation} \beta \sim \langle \exp(-\tau) \rangle = \frac{1}{\tau} \int_{0}^{\tau} \exp(-\tau') d\tau' =  \frac{1-\exp(-\tau)}{\tau} \text{,} \end{equation} which coincides with the expression for a radially expanding sphere.

\begin{figure*}
	\plotone{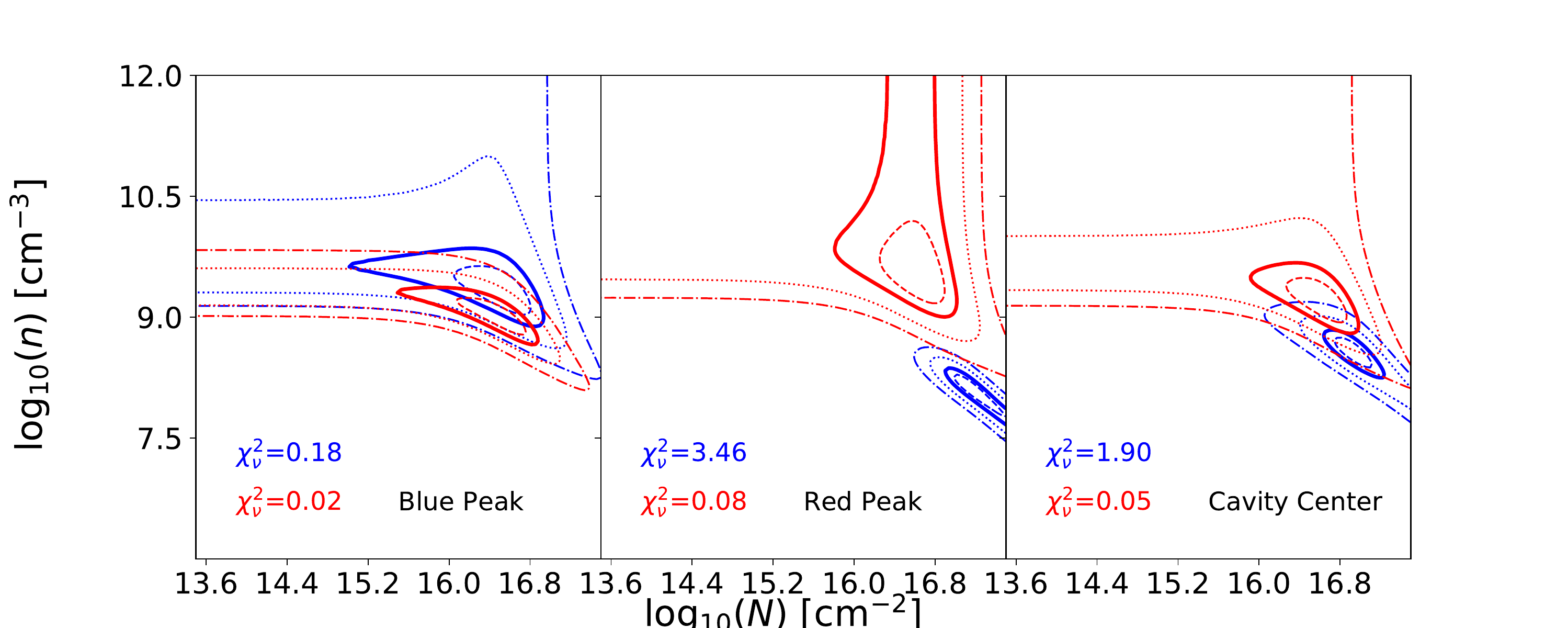}
	\caption{$\chi^2$ map in the density - column density parameter space. The dashed, solid, dotted and dashed-dotted lines represent the 0.5, 1, 2 and 3$\sigma$ confidence intervals. The red and blue colors represent the fits for the red and blue wings emission, respectively. The labels for each position are listed in Table~\ref{table:all_points}. \label{fig:chi2}}
\end{figure*}

The procedure consists of using the integrated intensity ratio of several SO$_2$ lines to constrain the physical parameters \citep[e.g.][]{fu_2012}. Since \emph{RADEX} does not include data on every SO$_2$ transition, and also some lines of SO$_2$ cannot be used (they are blended with neighboring lines in the blue and/or red wing), only 4 lines were used for this analysis. The lines should be as far away as possible in terms of energy, for the same reason that in the rotational diagram case. Two of them have low and two of them have high upper-level energies ($E_u$), hence 4 line ratios are defined: (J$=20_{0,20}-19_{1,19}$)/(J$=8_{4,4}-8_{3,5}$), (J$=20_{0,20}-19_{1,19}$)/(J$=11_{4,8}-11_{3,9}$), (J$=25_{3,23}-25_{2,24}$)/(J$=8_{4,4}-8_{3,5}$) and (J$=25_{3,23}-25_{2,24}$)/(J$=11_{4,8}-11_{3,9}$).

The code needs several input values. Density and column density are left as free parameters. The kinetic temperature is fixed to $170$\,K, which is the value constrained by SO$_2$. The line-width is set to $16.5$\,km\,s$^{-1}$, which is the average of the FWHM of the 4 lines. The background temperature is set to $2.73$\,K. The main collision partner that \emph{RADEX} handles is H$_2$. The output of the code for each pair of density and column density values are the line temperature $T_R$, the excitation temperature $T_{\rm ex}$, the optical depth $\tau$ of the line and the integrated intensity assuming a Gaussian shape. The latter is used to calculate the line intensity ratio.

Our method is similar to the one used by \citet{plume_1997} and \citet{vandertak_2000}. It consists of the following: given $N$ line ratios (in this case $N=4$) and given the \emph{RADEX} model that will predict line ratios as a function of density and column density, the quantity that must be minimized is
\begin{equation} 
\chi^2 = \sum_i \frac{(R_i - R_{m,i}(n({\rm H_2}),N({\rm SO_2})))^2}{\sigma_i^2} \text{,} 
\end{equation}
where $i$ stands for each of the defined 4 ratios, $R_i$ is the observed line ratio, $R_{m,i}$ is the \emph{RADEX} modeled line ratio and $\sigma_i$ is an estimation of the error in the line ratio. 

The wing emission in the SO$_2$ lines must be used to measure the line ratios. These ranges were set as $-120.0$ to $-95.6$\,km\,s$^{-1}$ for the blue wing and $-79.9$ to $-50$\,km\,s$^{-1}$ for the red wing. These limits are slightly different than the ones defined in Section~\ref{sec:rot_dia}, since we are considering a subset of SO$_2$ lines. We estimate the $\sigma_i$ errors in the following way: the error due to the RMS of the spectra is $\sim 1$\%. The calibration error (phase and bandpass) is $\sim 15$\%. Considering an error of 15\% in the integrated intensity, the resulting error in line ratios is $\sim 20$\%, which will be the adopted value. For each location, the averaged spectra within a circle of $1\farcs0$ of diameter was used, which corresponds to roughly the size of one synthesized beam. Three relevant locations were chosen: the center of the cavity, as well as the red and blue peaks defined by the SiO emission. Their coordinates are listed in Table~\ref{table:all_points}.

Using the standard interpretation of the $\chi^2 = \chi^2(n_{\rm H_2},N_{\rm SO_2})$ space, the confidence levels are estimated as contours at chosen $\sigma$. The results for each of the locations probed are shown in Figure~\ref{fig:chi2}. The contours corresponding to 0.5, 1, 2, and 3$\sigma$ are shown as dashed, solid, dotted, and dashed-dotted lines, respectively. The results are tabulated in Table~\ref{table:all_points}. The blue and red peak conditions are representative of the conditions in the outflow. The measured density is quite high ($n \gtrsim 10^9$\,cm$^{-3}$). These conditions are representative of the dense and shocked gas in the outflow, as traced by SO$_2$.

\begin{figure}
	\plotone{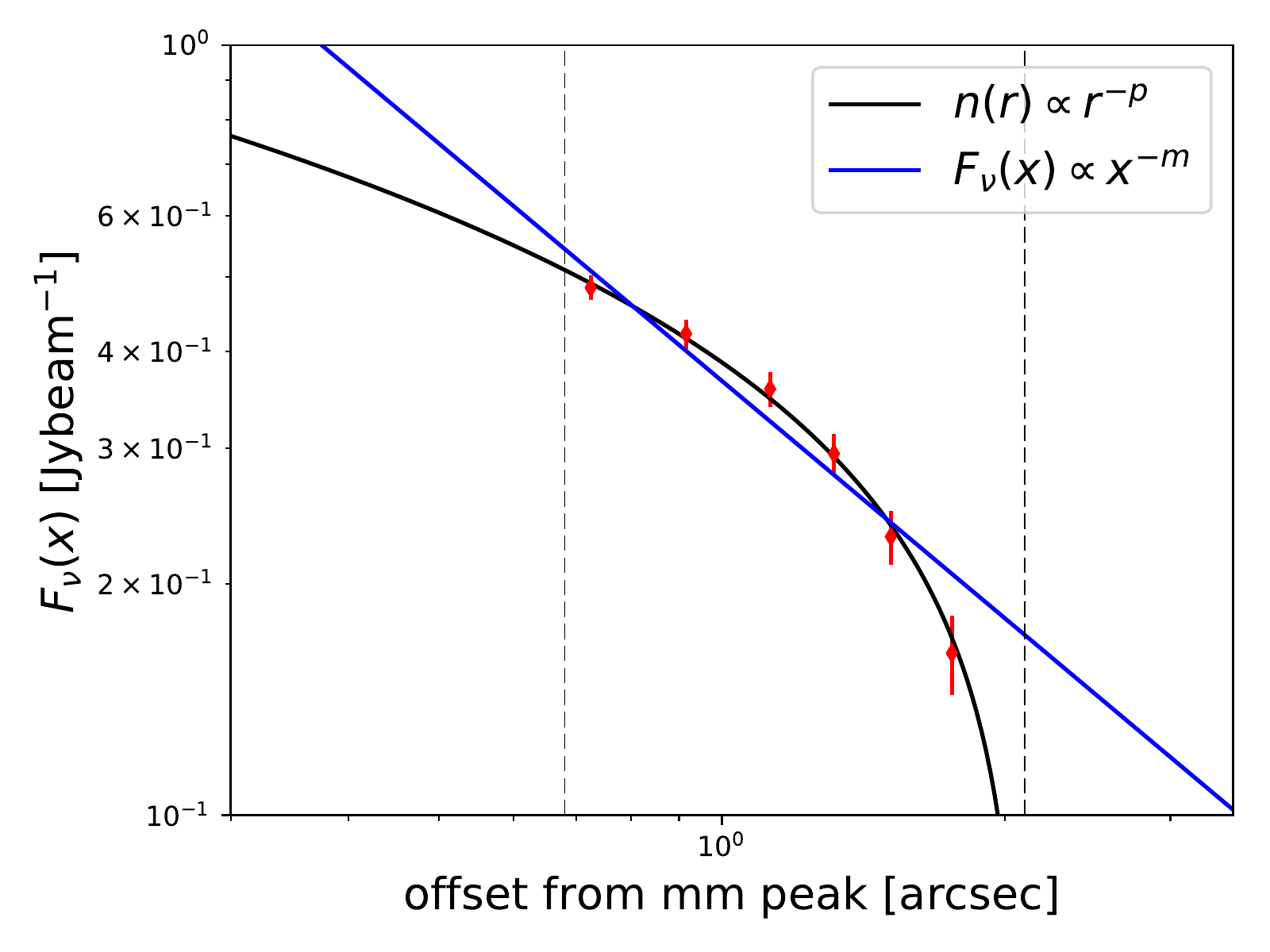}
	\caption{Diamonds: ALMA 862\,$\mu$m continuum radial profile measured with respect to the peak. The two vertical lines are the beam size and the $5\sigma$ contour radius from the map (the radius $R$ of the source). Blue line: Fit of a power law to the flux radial profile. Only the points inside the two vertical lines were considered. Black line: Flux radial profile calculated using an isothermal gas sphere with a density power law. See Section~\ref{sec:density_gradient} for details.\label{fig:radial_profile}}
\end{figure}

\subsubsection{Density radial gradient, dust continuum observations\label{sec:density_gradient}} One way to estimate the density distribution of the proto-stellar core is to use the thermal dust continuum observations: in our case, a line-free band in our ALMA data set. In the literature, this is usually performed assuming that the density and temperature will follow a power law with the radius, i.e. $n \propto r^{-p}$ and $T \propto r^{-q}$. For example, in \citet{garay_2007}, 19 IRAS point sources were imaged with the SEST telescope in 1.2\,mm. They find that most of these sources are single peaked, with a density power-law index in the range $p \sim 1.5-2.2$, which means that massive dense cores can be described as centrally condensed. In \citet{beuther_2002_cores}, they characterize 69 massive star forming regions using the 1.2\,mm continuum emission. They conclude that a unique power-law is not enough to characterize the radial distribution, but a flat center, followed by an inner and an outer radial power-law is adequate. The mean value for the inner power-law they found is $p \sim 1.6 \pm 0.5$. However, all of these studies are using single-dish observations and are appropriate on clump scales of $\sim 0.25$\,pc. The theory says that assuming a density power-law $n \propto r^{-p}$, a temperature distribution $T \propto r^{-q}$ and an observed flux distribution $F \propto r^{-m}$, in the optically thin emission, the three coefficients are related by $m = p + Qq - 1$ \citep{adams_1991}, where $Q$ is a frequency correction factor ($\sim 1.1$ for 0.87\,mm continuum). In a related method, \citep[e.g. ][]{looney_2003}, the density power law is directly estimated from the interferometric visibility space, rather than fitting in the deconvolved flux (mimicking what it is performed on single-dish flux observations).

Using the 862\,$\mu$m continuum ALMA observations, a radial profile is calculated using contours of equal flux. Given two contiguous levels, all the pixels in between the two levels are averaged. This flux is associated to a radius $(b_i + b_{i+1})/2$ where $b_i = \sqrt{A_i/\pi}$, i.e., the average of the two radii associated to the effective area between contour $i$ and $i+1$. As external radius of the clump, we use $2\farcs1$, which is the radius of the $5\sigma$ contour level in the 862\,$\mu$m emission. The radial profile is presented in Figure~\ref{fig:radial_profile}. Assuming a density function $n(r)$, the column density in a sphere can be calculated straightforwardly with the following equation \citep{dapp_2009} 
\begin{equation}
	N(x) = 2 \int_x^R \frac{n(r) r dr}{\sqrt{r^2 - x^2 }} \text{,}
\end{equation}
where $x$ is the angle distance with respect to the peak (the impact parameter), $r$ is the radial distance with respect to the center and $R$ is the sphere radius. In the optically thin limit, the flux is given by \citep{kauffmann_2008}
\begin{equation} \label{eq:flux_column_density}
	F_{\nu}(x) = B_{\nu}(T_{\rm d}(x))\kappa_{\nu} \mu m_H N(x) \text{,} 
\end{equation} 
where $B_{\nu}$ is the Planck spectral law and $T_{\rm d}(x)$ is the dust temperature.

A zeroth-order approximation is to consider the dust temperature of the core to be constant. On the one hand, our estimation of temperature comes from the rotational diagram of CH$_3$CCH, which indicates that the core has a temperature around 70\,K. On the other hand, a direct fit to the observed flux radial profile with a power-law gives an index of $m=1.0\pm0.2$. If we consider $T_{\rm d} = 70$\,K in equation~\ref{eq:flux_column_density}, we can fit a density power-law with an index $p=0.8\pm0.2$.

We only have one image of the dust continuum and we do not have an independent estimate for the dust temperature profile. Assuming a constant temperature is not ideal, as we have done in the previous paragraph. We can perform an improvement as a first approximation to consider a temperature profile, assuming that the mean dust temperature of the core is $\sim 70$\,K and looking into the literature for massive star-forming cores profiles. We perform the following steps:
\begin{enumerate}
	\item We assume that the temperature profile follows a power-law with index $q=0.4$, value used for high-mass star forming cores in the literature \citep[e.g.][]{vandertak_2000,garay_2010}.
	\item We assume that the mean temperature is 70\,K between the radii of $0\farcs7$ and $2\farcs1$. This is
	\begin{equation}
	    \langle T_{\rm d} \rangle = 70{\rm K} = \frac{1}{2\farcs1 - 0\farcs7} \int_{r=0\farcs7}^{r=2\farcs1} A r^{-0.4} dr \text{.}
	\end{equation}
	\item Under these conditions, the dust temperature profile is given by $T(x) \sim 58{\rm K}(x/2\farcs1)^{-0.4}$.
\end{enumerate}
In this more realistic approach, the fitted density model is a power-law with index $p=0.6\pm0.1$. This measured density power law is less steep than values reported for massive dense clumps, which could be due to the outflow moving gas outwards, therefore flattening the density profile. We note the lack of large scale emission by the interferometric observations (missing scales above $11.0-14\farcs0$)  and the flattening at small scales towards the center.


The fitted dust density power law is $n(r) = 1.8\times10^{4} (r/2\farcs0)^{-0.6} $\,cm$^{-3}$, so assuming a gas-to-dust ratio of 100, the mean gas density is $\sim 4.5\times10^{6}$\,cm$^{-3}$, which agrees with the high-density environments reported for massive star formation sources in the literature \citep[$>10^5$ cm$^{-3}$ in small scales $\lesssim 0.05$\,pc, ][]{evans_1999}. This density is $\sim 3$ orders of magnitude smaller than the constraint from Section~\ref{sec:radex}. However, the estimate from the 862\,$\mu$m thermal dust continuum traces all the warm gas coupled with the dust. We would expect this to be less dense than the shocked and compressed outflow gas.

\subsubsection{SiO shocks: model and properties} \label{sec:sio_shock} Following the model of \citet{gusdorf_2008}, we compare our observations of SiO(J$=8-7$) with the rotational spectra simulated in that work. They give a grid of models that allows to constrain the properties of the shock and outflow.

In this model, the release of SiO and related molecules is through the erosion of charged dust grains and their ice mantles by collision with neutral particles driven by a steady-state C-type shock. They consider multiple variables, such as the dynamics of the dust grains, the accurate description of the sputtering, the thermal balance, and a complex gas and solid phase chemical network. Finally, the intensities of the emission of the SiO spectrum are calculated through a LVG code (Section~\ref{sec:radex}).

\citet{leurini_2013} applies this model to the MYSO IRAS 17233-3606, \citet{klaassen_2006} uses the related model of \citet{schilke_1997} for the G5.89-0.39 outflow. They conclude that the SiO rotational emission spectra can be modeled in shocks located in high-mass star forming regions, using the tools developed for more quiescent low-mass star forming regions, but using higher values for pre-shock densities. \citet{gusdorf_2016} uses the shock model for Cepheus A, a massive star nursery, but for CO and OH line observations.

The parameters of the model that can be constrained and affect the model significantly are the pre-shock density $n_{\rm H}$ and the shock velocity $v_{\rm s}$. The transverse magnetic field and the viewing angle are also considered in the model, but found to have a minor influence.

For this analysis, we use the single-beam ($\sim 18\farcs0$) observations of SiO(J$=7-6$) and (J$=8-7$) from \citetalias{merello_2013} obtained with APEX. According to \citetalias{merello_2013_alma}, the differences between the single APEX spectrum and the integrated ALMA spectrum of SiO(J$=8-7$) are less than 10\%. The (J$=7-6$)/(J$=8-7$) line ratio from the APEX data is $0.86 \pm 0.02 $ (red wing) and $0.89 \pm 0.03$ (blue wing). The best model of shock that fits this is from Figure 9 in \cite{gusdorf_2008}. Noting that most shock models have line intensity ratios decreasing at high $J_{\rm up}$, only two models can give an SiO(J$=8-7$) intensity greater than the SiO(J$=7-6$) one. Both have $n_H=10^6$\,cm$^{-3}$ and $v_s=32$ and $34$\,km\,s$^{-1}$. The modeled line ratios are $\sim 0.93$ and $\sim 0.88$, respectively. We conclude that the shocks traced by the SiO observations have a speed of $v_s \sim 34$\,km\,s$^{-1}$ and pre-shock densities of $n_H \sim 10^6$\,cm$^{-3}$. Similar values are found by other works in massive outflows \citep{leurini_2013,leurini_2014,gusdorf_2016}.

\subsection{Geometry and kinematics}

\begin{figure}
	\plotone{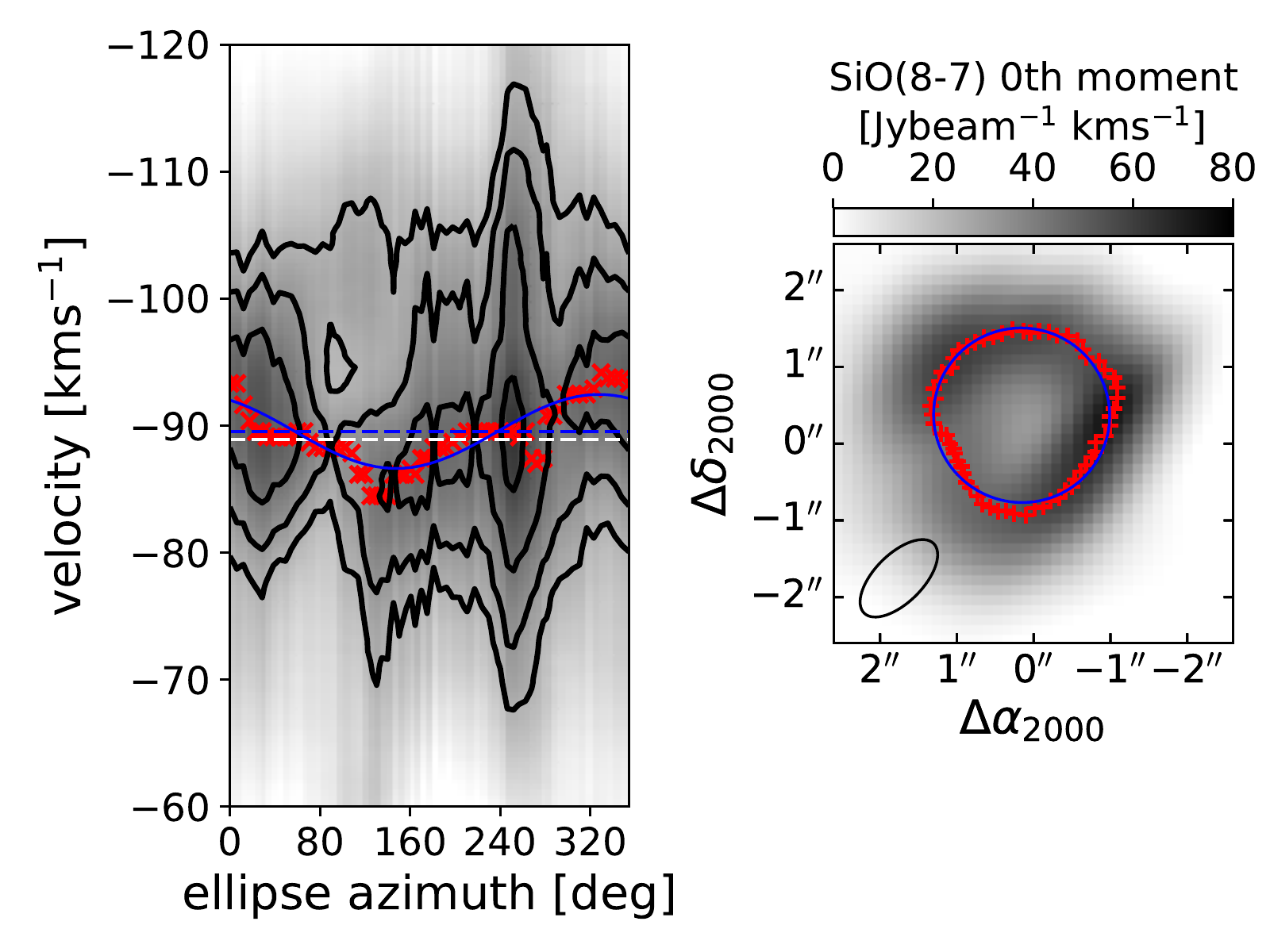}
	\caption{Left: On grayscale, the azimuthal PV plot of the SiO(J$=8-7$) line. The red crosses represent the velocities where the peak emission is for each ellipse azimuth. The blue horizontal line is the fitted $V_0$ and the white line is the systemic velocity $-88.9$\,km\,s$^{-1}$. The blue line corresponds to the best fitted $V_Z(\theta)$. Right: On grayscale, the SiO(J$=8-7$) 0th moment map at ambient velocities. The red crosses trace the observed ring. The blue line is the best fitted projected ellipse. \label{fig:fit_anillo_sio}}
\end{figure}

\begin{figure*}
	\plotone{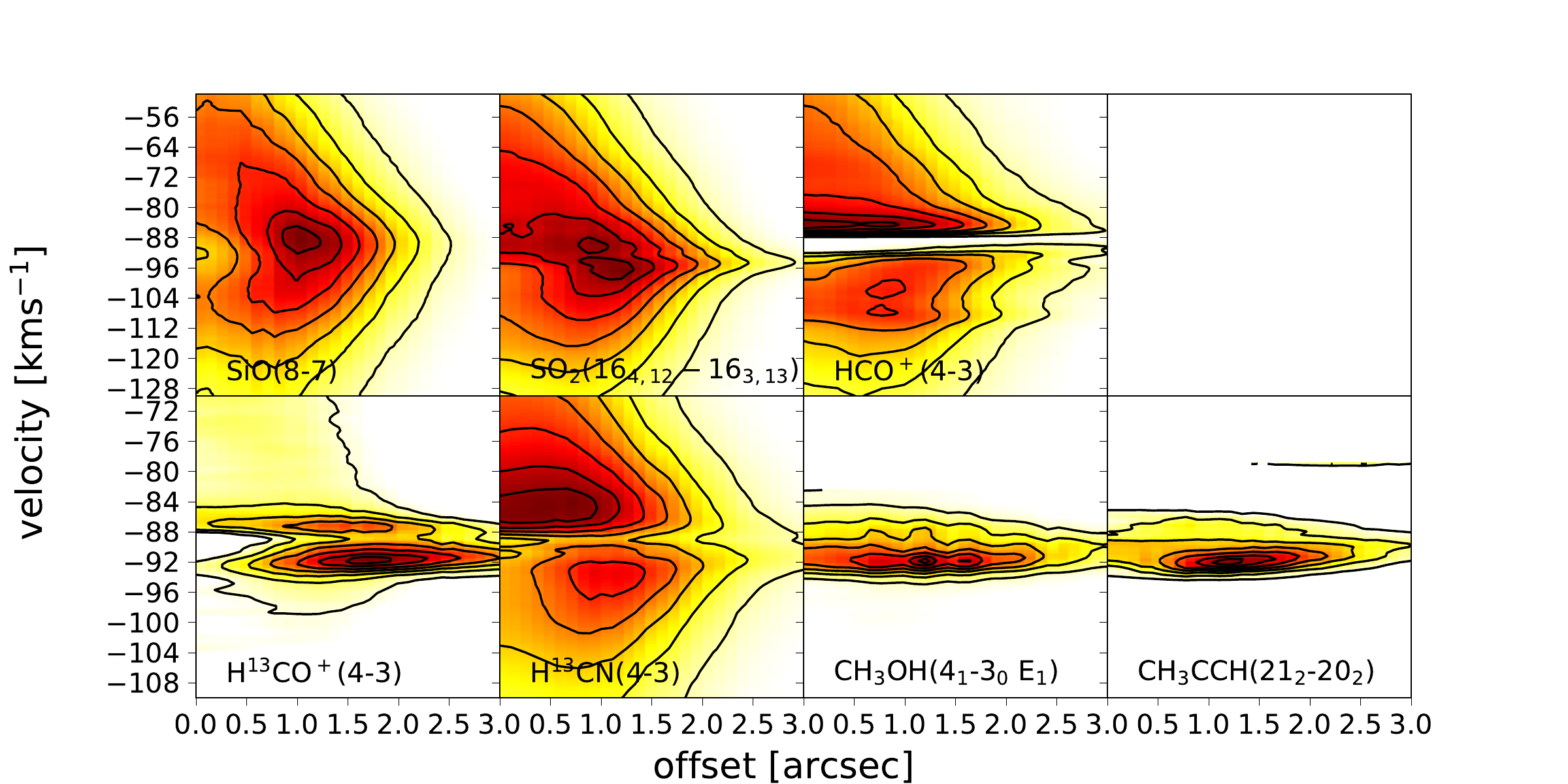}
	\caption{Radial direction PV plots for several observed lines. The contours are at 10, 25, 40, 55, 70, 85 and 95\% of the peak value. The center of the cavity (the 0 coordinate of the X-axis) is $\alpha_{J2000}=16^h12^m10^s.0$, $\delta_{J2000}=-51^{\circ}28'37\farcs45$. \label{fig:obs_radial}}
\end{figure*}

\subsubsection{Radial and Azimuthal PV plots} PV plots along a straight line axis can give information only along that particular axis, so it is helpful to plot the same information in a different fashion. We use the two natural directions of polar coordinates: radial and azimuthal. The radial PV plot is constructed by choosing a center point, and the pixels at the same radius within a ring are averaged, using the function \emph{kshell}, part of the KARMA package \citep{karma_paper}. The azimuthal PV plot is constructed by fitting an ellipse to trace the ring-like emission in the integrated intensity map. At each azimuthal angle around the perimeter of the ellipse, we average in a slit of 11 pixels, perpendicular to the tangent of the ellipse \citep{lopez_2016}. For a full description, see Section~\ref{sec:ring_model}.

An expanding shell geometry should show an ``inverted C'' pattern in the radial PV plot \citep{purcell_2009}. At the center of the shell, there is little emission at the systemic velocity, and at $\pm$ the expansion velocity we see the emission. The radial PV plots for several observed lines are shown in Figure~\ref{fig:obs_radial}. In this figure, the SiO panel presents the described ``inverted C'' profile, which hints an expanding shell. The shape is not sharply defined though, so estimating the expansion velocity only from this plot is difficult. The peak of emission is clearly defined, peaking at a radius of $\sim 1\farcs0$ ($\sim 0.036$\,pc at the source distance). The SO$_2$ and HCO$^+$ lines have a very similar radial PV plot, however the latter is completely self-absorbed, and we cannot draw further conclusions. H$^{13}$CO$^+$, CH$_3$OH and CH$_3$CCH show the systemic velocity emission, but the first also shows weak high-velocity wings. The H$^{13}$CN and H$^{13}$CO$^+$ lines show a small self absorption dip at the systemic velocity. The CH$_3$OH and the CH$_3$CCH lines have a single peak of emission, located at a radius of $\sim 1\farcs2$ ($\sim 0.044$\,pc). In H$^{13}$CO$^+$, the emission peaks at $\sim 1\farcs7$ ($\sim 0.062$\,pc), which makes H$^{13}$CO$^+$ the line that traces the most extended emission. 

\begin{figure*}
	\plotone{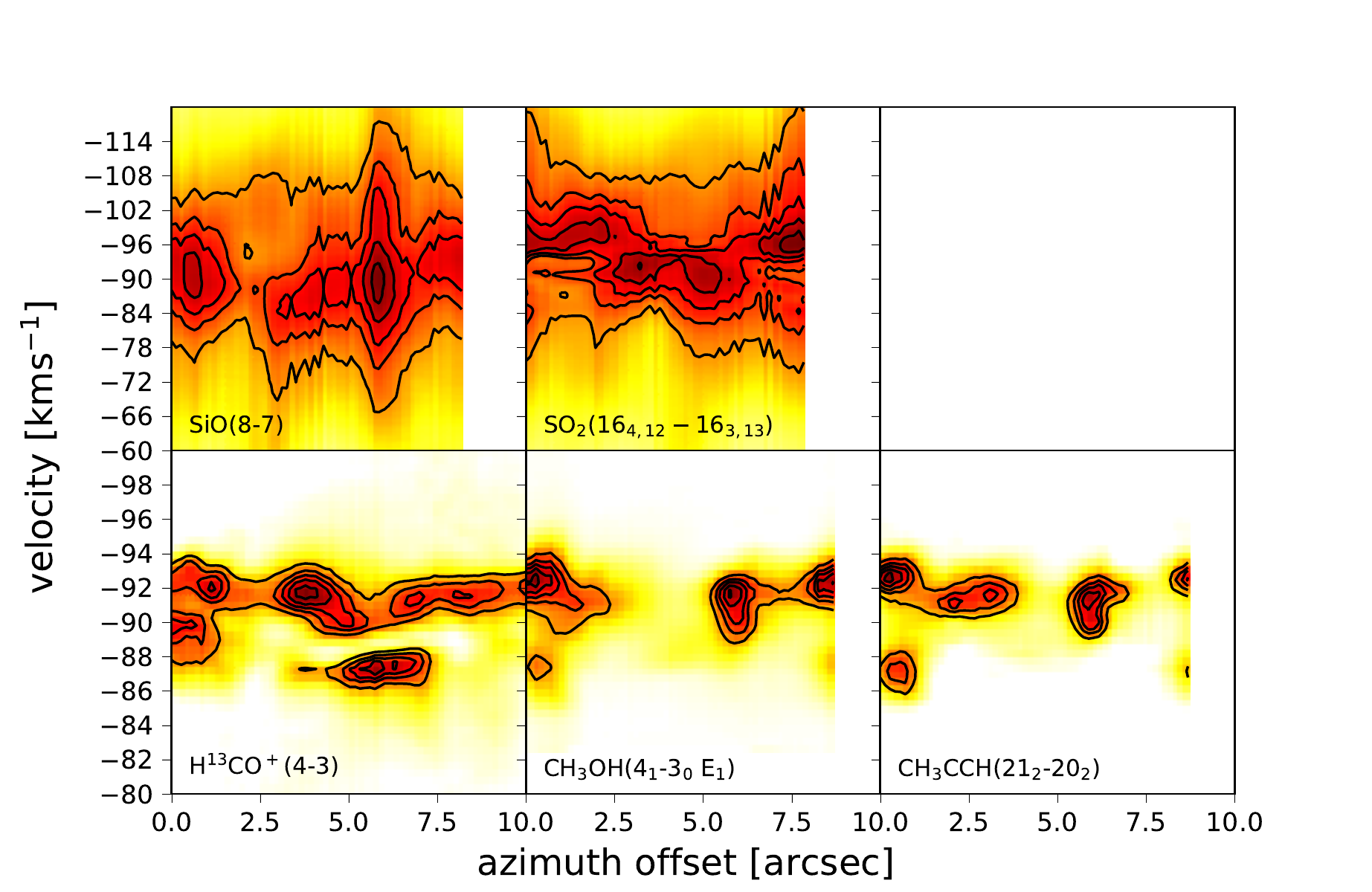}
	\caption{Azimuthal direction PV plots for several observed lines. The contours are at 50, 65, 75, 85 and 95\% of the peak value.\label{fig:obs_azimuth}}
\end{figure*}

An expanding perfect sphere appears isotropic in an azimuthal PV plot. However, a shell, which in principle can have an ellipsoidal shape, should be sinusoidal in the azimuthal PV plot \citep[e.g. ][]{lopez_2016}. For the same lines shown in Figure~\ref{fig:obs_radial}, the azimuthal PV plots are shown in Figure~\ref{fig:obs_azimuth}. The SiO line can be more readily interpreted. Around $-90$\,km\,s$^{-1}$, the sinusoidal oscillation can be seen in the peaks as a function of azimuth angle. In Section~\ref{sec:ring_model}, an analytical ring model is fitted to this sinusoidal oscillation, shown in Figure~\ref{fig:fit_anillo_sio}. The SO$_2$ azimuthal plot has a similar shape. However, in the latter case, the oscillation is at an offset velocity with respect to the SiO. A sinusoidal shape could be invoked, but the fact that it is centered at a slightly bluer velocity than the systemic $-90$\,km\,s$^{-1}$ makes this plot harder to interpret. The H$^{13}$CO$^+$, CH$_3$CCH and CH$_3$OH lines are very similar, showing clumpiness at the systemic velocity.

\subsubsection{Ring model for the SiO(J$=8-7$) azimuthal PV plot} \label{sec:ring_model} Figure~\ref{fig:obs_azimuth} shows that the azimuthal PV emission of the SiO line seems to trace some sort of expanding motion, as mentioned above. Using the best-fit ellipse points, together with the azimuthal PV plot, a simple mathematical model of a ring is fitted. When referring to ring, we mean the ring-like emission seen in SiO. The method follows from \citet{lopez_2016}.

Consider a ring in the plane XY of the sky expanding in the outwards direction at a constant velocity. The equations that describe each XY coordinate and the projected velocity in the Z direction perpendicular to the plane of the sky are
\begin{eqnarray}
	X(\theta) =& X_0 + R(\cos\theta \cos\alpha \cos\beta - \sin\theta \sin\beta ) \label{eq:anillo1} \\
	Y(\theta) =& Y_0 + R(\cos\theta \cos\alpha \sin\beta + \sin\theta \cos\beta) \label{eq:anillo2}\\
	V_z(\theta) =& V_0 + V\cos\theta \sin\alpha \label{eq:anillo3} \text{,}
\end{eqnarray}
where $\theta$ is the angle along the perimeter of the ellipse (the azimuth angle), $R$ is the radius of the ring, $V_0$ is the systemic velocity, $V$ is the constant expansion velocity of the ring, $\alpha$ is the inclination respect to the plane of the sky, and $\beta$ is the position angle (rotation of the XY plane). Using the best-fit ellipse, the observed $X$ and $Y$ coordinates are chosen for each ellipse azimuth angle as an intensity-weighted mean of the 10 pixels, belonging to the perpendicular ``slit'', that are considered when averaging the spectra. In this way, the best-fit ellipse acts as a guide for the ``real'' $XY$ points on the ring. In the same way, the $Z$ direction projected velocities are chosen as the velocity where the peak of the emission takes place for each ellipse azimuth angle, in the PV plot shown in Figure~\ref{fig:obs_azimuth}.

There is a list of observed coordinates $X$,$Y$ and $V_{Z}$ for each ellipse azimuth angle. Further, there is a list of $X$,$Y$ and $V_\text{Z}$ values, calculated with equations \ref{eq:anillo1}-\ref{eq:anillo3} and $\theta$ between $0$ and $2\pi$. A function calculating the sum of the distance in the 3D space ($X$,$Y$,$V_{Z}$) between the observed and modeled coordinates is defined. Then, this function is minimized for the best fit 7 parameters:  $X_0,Y_0,\alpha,\beta,R,V_0$ and $V$. The results are shown in Figure~\ref{fig:fit_anillo_sio}. On the left panel, the peaks at each corresponding ellipse azimuth are shown as red crosses. These trace the proposed inclined ring with their sinusoidal shape. The model of projected V$_{Z}$ velocity is shown as the solid blue line. On the right panel, the modeled projected ellipse in the plane of the sky is shown against the SiO 0th moment map at systemic velocities.

\begin{deluxetable}{c l}
	\tablecaption{Best fitted parameters for the ring model.\label{table:ring_best_parameters}}
	\tablehead{\colhead{Parameter} & \colhead{Value}}
	\startdata
	Center J2000 & 16:12:10.0 -51:28:37.4	\\
	Line of sight angle $\alpha$ & -8$^{\circ}$ \\
	Position angle $\beta$  & 72$^{\circ}$ \\
	Radius $R$  &  $1\farcs2$ \\
	Expansion velocity $V$ & 21 km\,s$^{-1}$ \\
	Systemic velocity $V_0$ & -90 km\,s$^{-1}$ \\
	\enddata
\end{deluxetable}

The best fit parameters for the model are listed in Table~\ref{table:ring_best_parameters}. The center of the ellipse naturally will coincide with the measured cavity center in \citetalias{merello_2013_alma}. The angle $\alpha$ is small, which supports the hypothesis that the cavity-outflow complex is closely aligned with the line of sight. The expansion velocity of $\sim 20$\,km\,s$^{-1}$ is very similar to the one determined by \citetalias{merello_2013_alma}, derived by only analyzing the position-velocity plot along or perpendicular to the outflow axis.

To measure the significance of the model, we performed a $\chi^2$-test. The estimation of the error is critical for this purpose, since the value of $\chi^2$ can change greatly depending on the error. The error is estimated from the Gaussian distribution of the measured velocities around the velocity where the peak is located. The method to estimate the error in the determination of the peak velocity is the following: given a level of noise in the measurement of flux (or temperature) of the spectrum, $\Delta T_{\rm RMS}$, we calculate the error as the width in velocity in the gaussian-fit maximum temperature to the gaussian-fit maximum temperature minus $\Delta T_{\rm RMS}$. That is, we look for ($v-v_0$) such that $T_0(1 - \exp(-(v-v_0)^2/2/\sigma_T^2) )= \Delta T_{\rm RMS}$, where $T_0$ and $\sigma_T$ are the maximum temperature and the standard deviation of the Gaussian fit, respectively. We set as the error for the $\chi^2$-test $\sigma_v = v - v_0$ for each spectrum at each azimuth angle. Then, we calculate a $\chi^2$ statistic only based on the projected velocity fit, given by $\chi^2 = \Sigma_i (\frac{V_Z(\theta) -V_Z^i}{\sigma_v})^2$, where $i$ runs through each of the offset angles in the perimeter of the ellipse and $V_Z(\theta)$ is given by eq.~\ref{eq:anillo3}. The resulting reduced-$\chi^2$ is $\chi_{\nu} = 0.42$, implying that the model might \emph{overfit} the data and the consideration of errors is conservative.

\subsection{Radiative transfer modeling of the SiO(J$=8-7$) line data cube} \label{sec:radiative-transfer-model} This section presents a radiative transfer model of the SiO(J$=8-7$) line emission using all the knowledge gathered in the two previous subsections. The parameters that we use as input in the model are listed in Table~\ref{table:summary}. This model is performed with the 3D radiative transfer code \emph{MOLLIE}. The full detailed description of how the model is generated is in Appendix~\ref{appendix:model}. Here, we limit ourselves to present only the results.

\begin{figure}
  \plotone{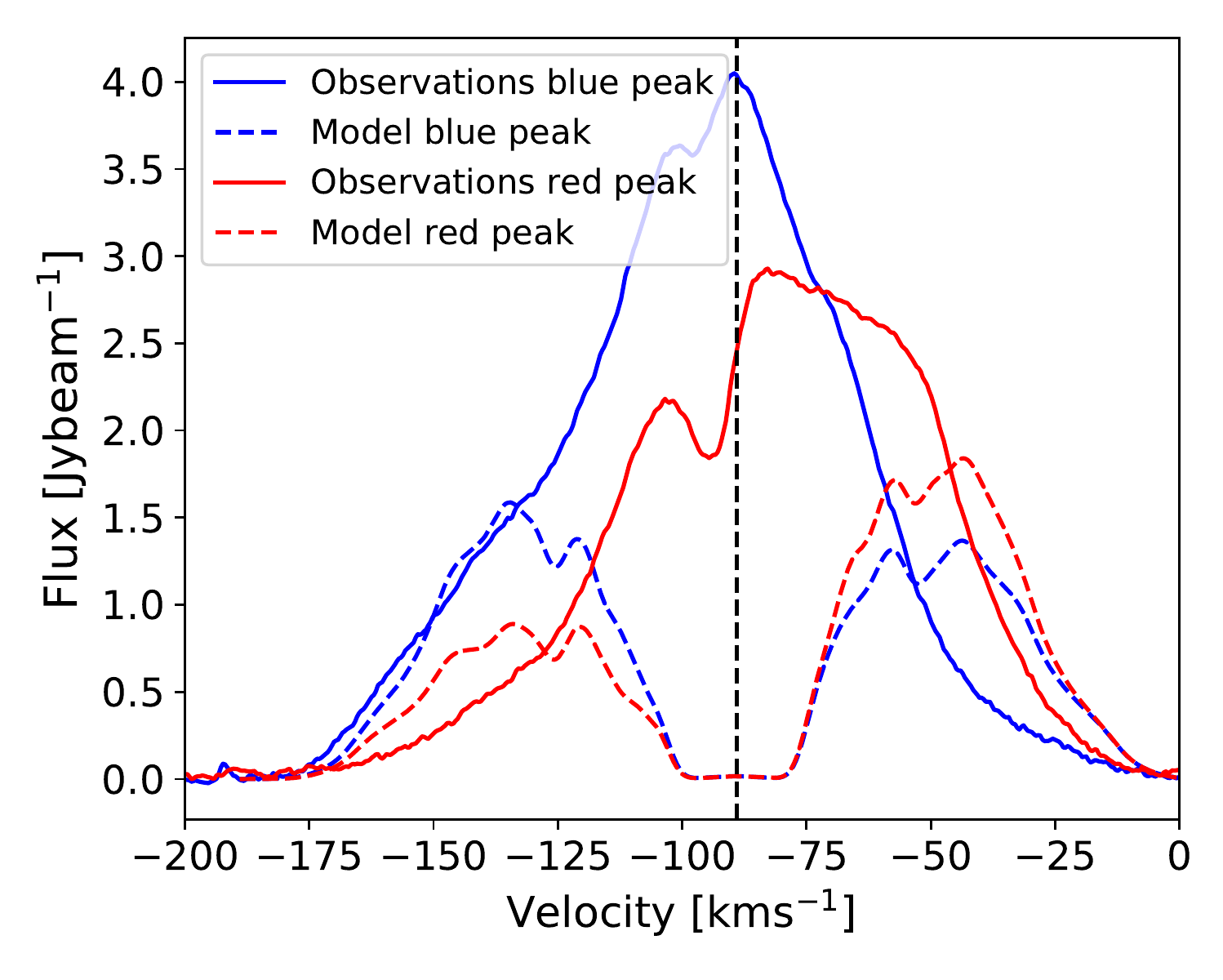}
  \caption{Comparison of averaged SiO(J$=8-7$) spectra, observations and outflow model. Each spectrum is averaged inside one beam. The blue and red color indicate spectrum on blue and red peaks ($\pm 40$\,km\,s$^{-1}$ channels), respectively. The solid lines represent observations. Dashed lines are for the MOLLIE model. The outflow model accounts for the high velocity channels $ | v | > 40$\,km\,s$^{-1}$. \label{fig:outflow_model}}
\end{figure}

We simulated the emission in two steps: one for an expanding shell, and one for a bi-conical outflow. The results of the outflow simulation are shown in Figure~\ref{fig:outflow_model}. We show the spectrum averaged over 1 beam ($1\farcs0$ diameter), for both observations and model, towards both the blue and red peaks. The wing emission is fairly matched by the model. While the high-velocity emission is present, as the velocity diminishes, the column of gas in the line of sight decreases (Figure~\ref{fig:outflow_model_profiles} in Appendix~\ref{appendix:model}). Therefore, the column of gas in the line of sight may be large, explaining the lack of flux at systemic velocities.

\begin{figure}
  \plotone{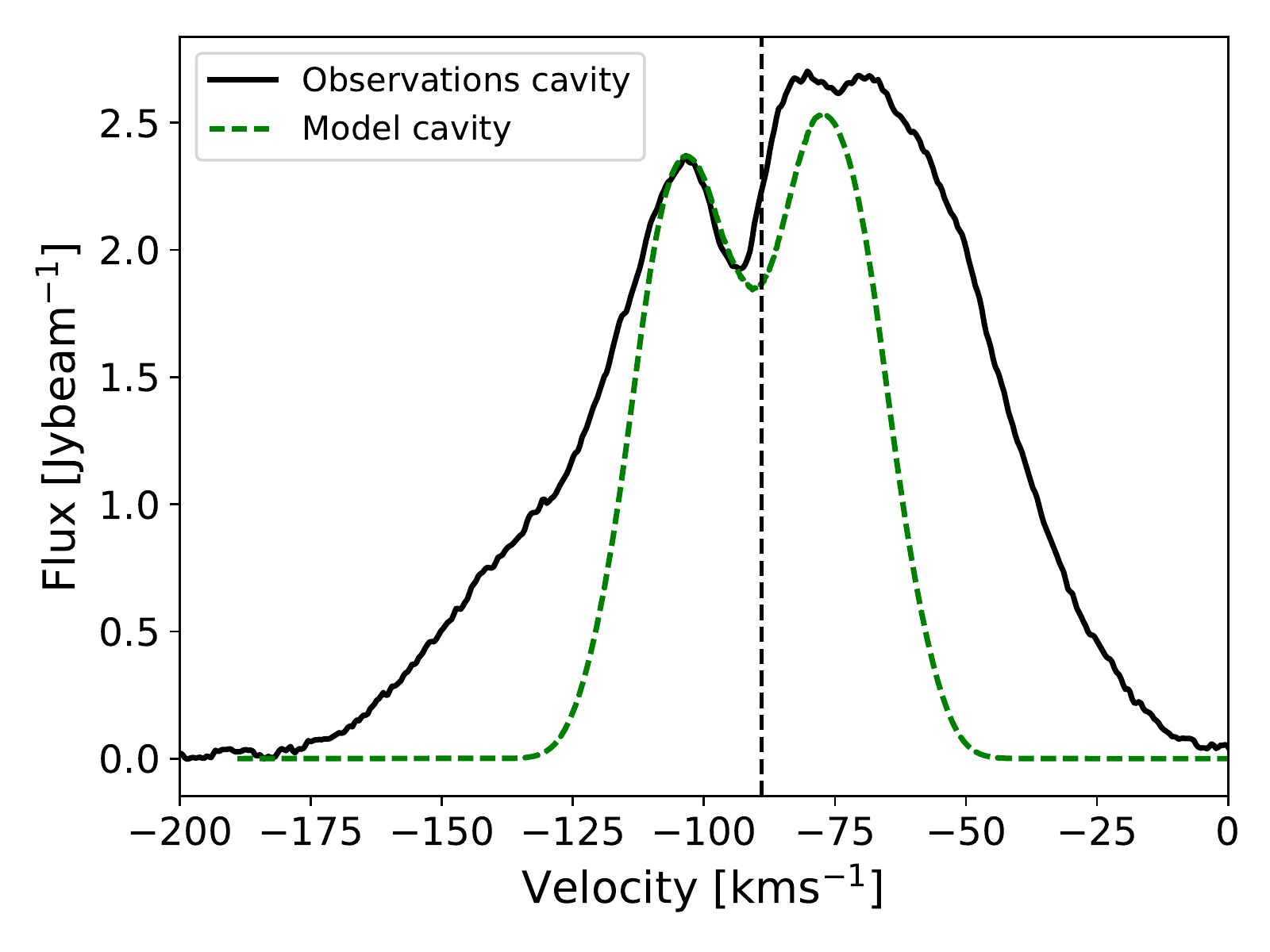}
  \caption{Comparison of observed and modeled SiO(J$=8-7$) spectrum inside the shell. The black solid line corresponds to the SiO(J$=8-7$) observed spectrum averaged over one beam inside the shell. The green dashed line is the modeled spectrum averaged over one beam inside the shell. The vertical line corresponds to the systemic velocity. \label{fig:cavity_model}}
\end{figure}

The averaged spectrum towards the SiO expanding shell is shown in Figure~\ref{fig:cavity_model}. The observed SiO profile of the shell exhibits an asymmetric profile, most likely evidence of the expanding motion. In this case, the red-shifted peak is stronger than the blue-shifted one, with a dip at the systemic velocity. However, it should be noted that the dip is not at the exact systemic velocity of $-89$\,km\,s$^{-1}$, but rather $\sim 3$\,km\,s$^{-1}$ blue-shifted. The modeled spectrum shows this same behavior, even though the intensity might not reproduce exactly the observed levels. It should also be noted that the red peak of the outflow is closer to the cavity center than the blue peak, therefore explaining the asymmetry between the blue and red wings in the spectrum in Figure~\ref{fig:cavity_model}.

\begin{figure*}
  \plottwo{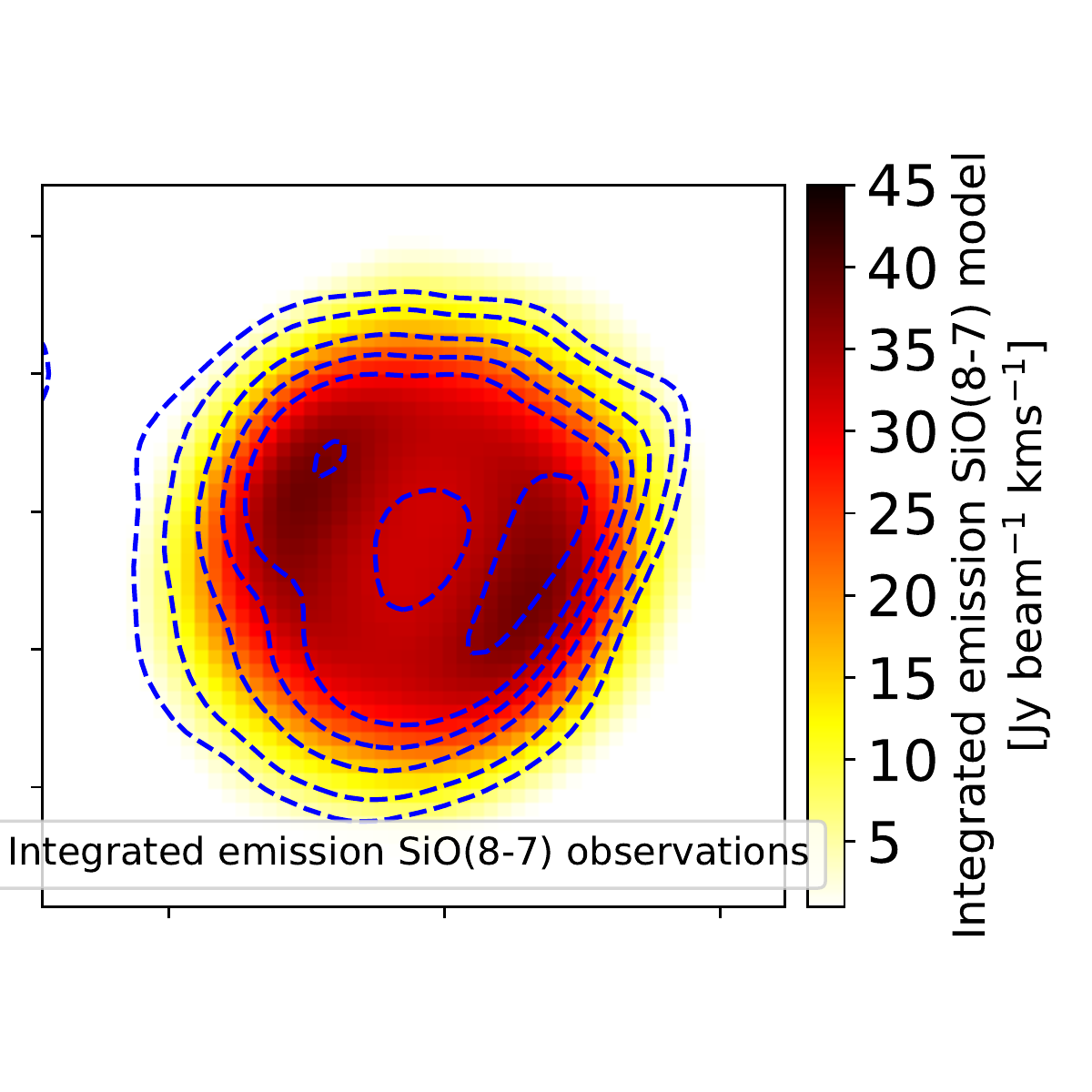}{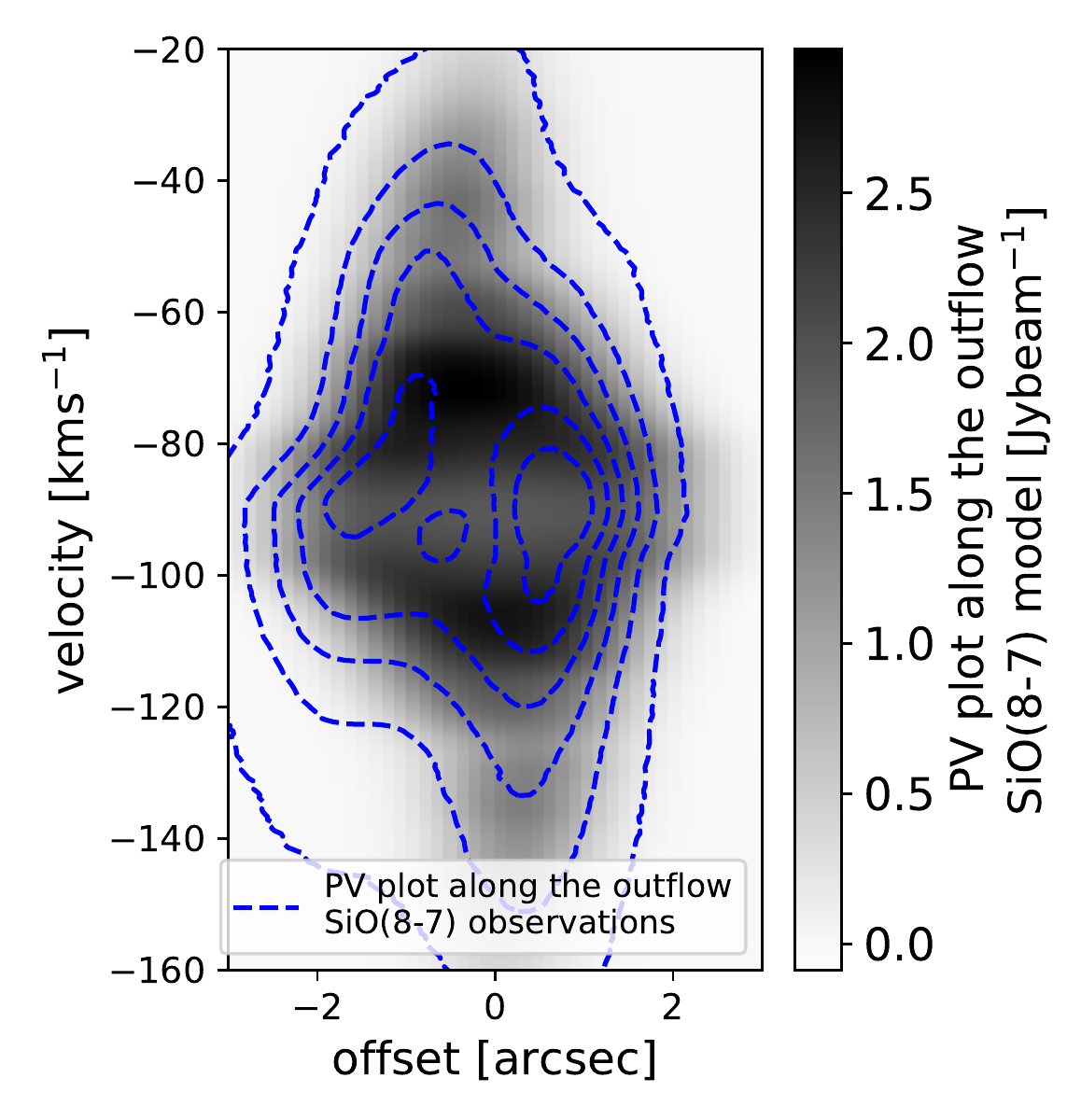}
  \caption{Left: Integrated emission at ambient velocity range of the simulated data cube, on color scale. The dashed blue contours correspond to the 0th moment map at the same range in the SiO(J$=8-7$) observations. The levels are 5, 10, 20, 30, 40, and 60 Jybeam$^{-1}$ km s$^{-1}$. Right: Position-Velocity plot along the outflow axis with a ``slit'' of $1\farcs54$. On gray scale, the modeled SiO(J$=8-7$) data cube. On dashed blue contours, the observations. The contour levels are 0.2, 0.86, 1.52, 2.18, 2.84, and 3.5 Jybeam$^{-1}$ \label{fig:total_model}}
\end{figure*}

In order to compare the emission of the SiO line observations with the modeled data cube, we add up both the shell and outflow models, since they are at non-overlapping velocities. However, we are ignoring any possible radiative interaction between the outflow and the shell. We compare with the observations using an integrated emission map at systemic velocities and the position-velocity plot along the outflow axis. Figure~\ref{fig:total_model} shows both of these. In general terms, the model is able to qualitatively reproduce the observations. Therefore, we can state that the following is consistent with the G331.512-0.103 massive outflow:

\begin{enumerate}
	\item A bipolar outflow with an opening angle of $\sim 30^{\circ}$, a maximum velocity around $85$\,km\,s$^{-1}$ at the axis that decreases with the cylindrical radius, as stated in \citet{stahler_1994}; and with an inclination of $\sim 8-10^{\circ}$ with respect to the line of sight. The interior of the outflow has low density ($\sim 10^2$ cm$^{-3}$) and high velocity. The outer layer of the outflow, that is in contact with the ambient core gas, has high density ($\sim 10^{5-6}$ cm$^{-3}$) and low velocity.
	\item A cavity, with low density $\sim 10^2$ cm$^{-3}$ and medium-velocity ($\sim 20$ km\,s$^{-1}$), with a radially outwards expansion. This structure has a radius of $\sim 2\farcs5$.
\end{enumerate}

However, there are some aspects of the observations that were not reproduced.

\begin{enumerate}
	\item The emission of SiO$(8-7)$ in the position of the ring-like emission at the systemic velocities is clearly stronger than the emission in the cavity center. This translates to a spectrum with a single peak, rather than a dip, along the ring-like emission. The model is not able to reproduce this, since the dynamics of the expanding motion is present in the entire central region of emission, hence a double peak spectrum with a dip at systemic velocity is present in the entire emission region of the model, both in the cavity and in the ring-like emission perimeter.
	\item Peaks of emission and structures are present within the ring-like emission in the observations. For this, the observed integrated emission map at the systemic velocities shows a clear contrast between the ring-like emission and the cavity center, i.e. the integrated emission is about twice in the ring-like emission compared with the cavity center. However, the levels of emission inside the cavity center are similar in both the model and the observations.
\end{enumerate}

To improve the model, possible inhomogeneities present in the shell must be accounted for. The fact that the data show some level of clumpiness at the ring-like emission of SiO(J$=8-7$) indicates that the shell is not homogeneous and there is some structure beyond our simple analytical model.
\section{Summary} \label{sec:conclusions}

\begin{deluxetable}{l c}
\tablecaption{Summary of the properties derived for the source. \label{table:summary}}
\tablehead{ \colhead{Property}	& \colhead{Value}  }				
\startdata
Distance & 7.5 kpc \\
Mass of core & $\sim 40 M_{\odot}$ \\
Mass outflow lobes & $\sim 25 M_{\odot}$ each \\ 
Kinetic age & $\sim 2000$ yrs \\
Velocity range of the outflow & $\pm 70$ kms$^{-1}$ \\
Expansion velocity of cavity & $\sim 21$ kms$^{-1}$ \\
Mean density & $4.4 \times 10^6$ cm$^{-3}$ \\
Density power law index & $\sim 0.8$ \\
Temperature of core & $\sim 70$ K \\
Outflow inclination & $\sim 10^{\circ}$ 
\enddata
\end{deluxetable}

The G331.512-0.103 molecular core contains one of the most luminous and powerful massive outflows harbored in a high-mass star forming region in our Galaxy. We analyzed several molecular spectral lines observed at $\sim 1\farcs0$ resolution with ALMA band 7, deriving the physical conditions, morphology and kinematics of the source. Table~\ref{table:summary} summarizes the main derived properties of the source.

Based on the PV diagrams of the lines, there are two groups. One that traces the broad high-velocity wings, and one that traces narrow systemic velocity emission.

For the high-velocity outflow, we have determined its properties. The temperature of the most dense and hot gas in the outflow is $\sim 150$\,K (blue lobe) and $\sim 200$\,K (red lobe). The column densities of the SiO and SO$_2$ are constrained. These estimates are also supported by an analysis with the code RADEX: a density of $\sim 10^9$\,cm$^{-3}$ is constrained for the outflow shocked gas. The fractional abundance of SiO and SO$_2$ is in agreement with values found in other massive outflows in MYSOs.

The properties of the systemic velocity emission, the ambient core, are analyzed. The kinetic temperature of the ambient core is $T_{\rm K} \sim 70$\,K. The 862\,$\mu$m dust continuum emission can be well fitted with a density power-law with an index $\sim 0.8$ and a mean value of $\bar{n_{\rm H_2}}=4.4\times10^6$\,cm$^{-3}$.

The geometry and morphology of the ambient core is characterized by the peaks of the PV plots in the radial direction: structures with radii of $\sim 1\farcs0$ (SiO), $1\farcs2$ (CH$_3$OH) and $\sim 1\farcs7$ (H$^{13}$CO$^+$). A ring model was fitted to the SiO(J$=8-7$) azimuthal PV plot. The parameters of this model is an inclination angle of $\sim 10^{\circ}$ and an expansion velocity of $\sim 20$\,kms$^{-1}$.

To model the source, composed of an outflow and an expanding shell, we performed a radiative transfer model of the SiO(J$=8-7$) line using the code MOLLIE. The model is composed of two structures: a conical bipolar outflow with a velocity field that scales with the cylindrical radius, and an ellipsoidal cavity and expanding shell with a peak expansion velocity of $\sim 20$\,kms$^{-1}$ in the spherically radial direction. The model was able to reproduce the main features at the wings (outflow) and ambient velocity ranges, although the emission is not completely recovered. The strong peak that dominates at systemic velocities, observed in a ring-like emission surrounding a cavity, is not totally accounted for in the model.

Our global scenario for the source is the following: At the center position, where a newborn massive star is located, there are structures proper to the ambient core, which is at systemic velocities. The proto-star shows a massive, bipolar, and high-velocity outflow with velocities of $\pm 70$\,kms$^{-1}$, likely powered by a collimated jet and by the accretion of gas onto the source. The powerful stellar winds and ionizing radiation from the proto-star push against the ambient core gas, inflating a cavity and an expanding shell-like structure.

\begin{acknowledgements}
C.H.C. acknowledges support by CONICYT Beca de Magister Nacional, folio 221220026, and partial support by FONDECYT project 1120195. M.M. acknowledges support from the grant 2017/23708-0, S\~ao Paulo Research Foundation (FAPESP). L.B. and G.G. acknowledge support from CONICYT project Basal AFB-170002. We thank Al Wootten and the staff of NRAO for their help and assistance with the reduction of ALMA data. This Paper makes use of the following ALMA data: ADS/JAO.ALMA\#2011.0.00524.S. ALMA is a partnership of ESO (representing its member states), NSF (USA), and NINS (Japan), together with NRC (Canada) and NSC and ASIAA (Taiwan), in cooperation with the Republic of Chile. The Joint ALMA Observatory is operated by ESO, AUI/NRAO, and NAOJ.
\end{acknowledgements}

\bibliography{biblio.bib}

\appendix
\section{A: Integrated ALMA band 7 spectra of G331.512-0.103} \label{appendix:complete_band} 
\begin{turnpage}
\begin{figure*}
	\plotone{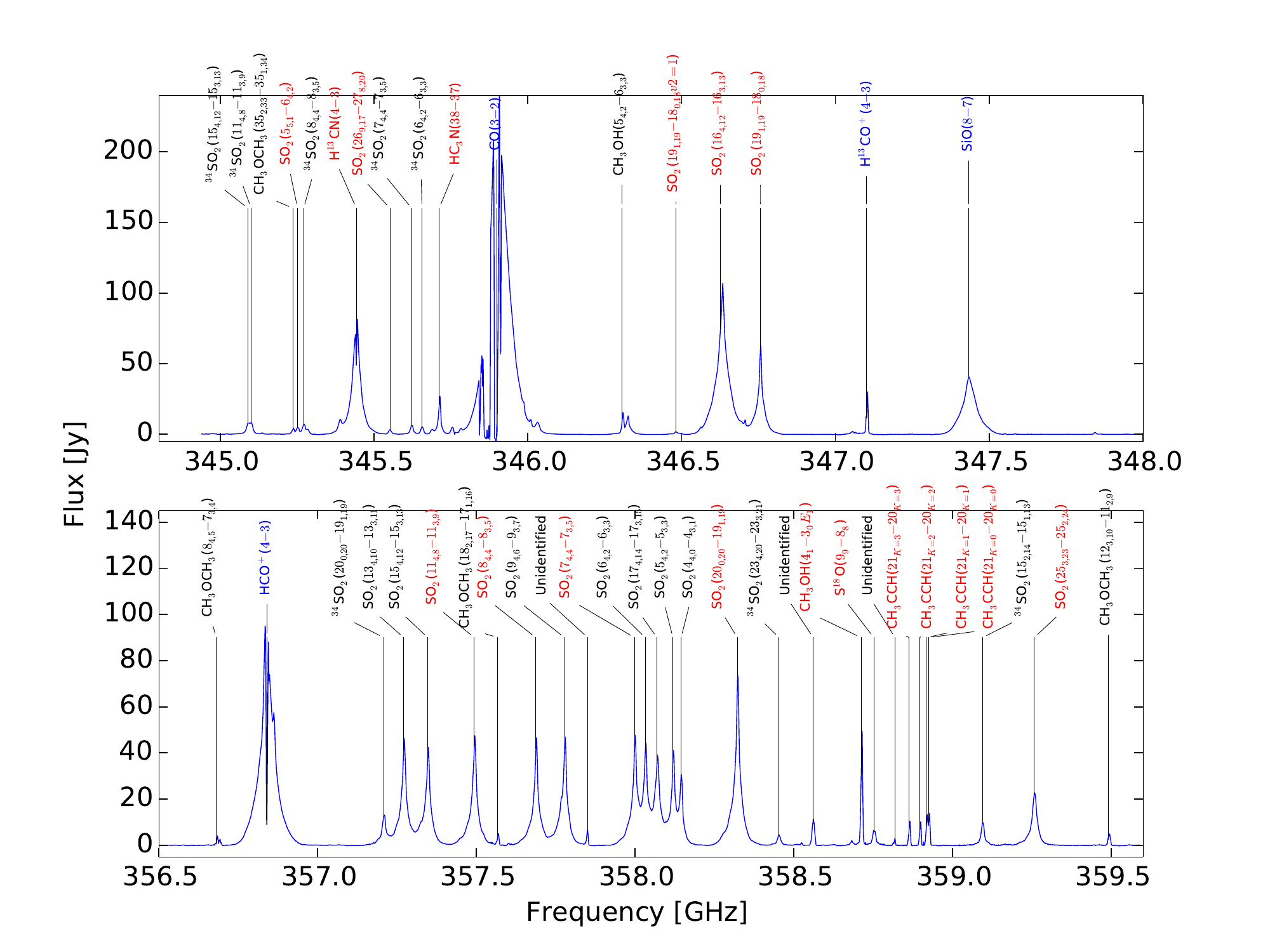}
	\caption{Composed SW3-SW2 (top) and SW0-SW1 (bottom) bands of the integrated spectra observed toward G331.512-0.103 observed with ALMA. The most prominent lines are marked across the spectra. Lines used in this work are colored. \label{fig:spectral_windows}}
\end{figure*}
\end{turnpage}

Figure \ref{fig:spectral_windows} shows the integrated spectra obtained with ALMA band 7 over the frequency range 345-348\,GHz (SW3-SW2), and 356.5-359.5\,GHz (SW0-SW1). Here we show all identified lines. The ones with blue labels correspond to the lines presented in \citetalias{merello_2013_alma}, the ones in red labels are the 18 lines newly analyzed in this study and the ones with black labels are the rest of the lines that will be used in future studies. 

\section{B: Detailed description of MOLLIE modeling of the source} \label{appendix:model} 

\subsection{Estimation of physical conditions} Since physical conditions can vary across the source, and multiple combination of geometries and parameters can reproduce an observed spectrum, we need a rough estimate of these conditions in order to constrain how the model is set up. In the following, we discuss broad estimates of the necessary physical conditions that must be specified in a MOLLIE grid model. For a 3D grid, we must specify for each voxel the density, the kinetic temperature, the local linewidth, the 3 components of the velocity vector and the fractional abundance of the molecule that is being observed.

The estimation of density is challenging. In Section~\ref{sec:physical_conditions}, we estimated the density of the ambient gas with the 862\,$\mu$m dust continuum and the density of the region emitting in SO$_2$, tracing the outflow wings and therefore the high-velocity and shocked gas. In the following models, the ``background'' density, i.e the density of the ambient core as a function of radius, is set to the power law estimated from the dust continuum flux in Section~\ref{sec:density_gradient},

\begin{equation}
	n(r) = 2.0 \times 10^6 [\text{cm}^{-3}] (r/1\farcs6)^{-0.8} \text{,} \label{eq:core_density}
\end{equation}

where $r$ is the radius expressed in arcsec. Note that this law corresponds to the ``background density'' in the model. It does not apply to the center of the model grid, where the shell or the outflow is to be modelled with a corresponding different density law. For the shocked high density, i.e. the density of the most dense section of the outflow and the cavity/shell (where the stellar winds from the proto-star are impacting), we will use the order of magnitude values derived in Section~\ref{sec:radex} for the blue and red peaks as a reference. That is $n = 10^9$ cm$^{-3}$.

We only have two estimates of temperature from 2 different molecules. One, the temperature estimated with the CH$_3$CCH molecule, tracing the core and systemic velocity environment. Since this molecule is a very good thermometer, we use it as the kinetic temperature of the ambient core. Then, we consider $T_K=70\pm7$\,K, the temperature of the main emission region. The other molecule that gives temperature estimates is SO$_2$. Since this molecule is likely tracing the most dense and shocked section of the gas, as stated before, we use it as the probe of outflow conditions. We use the value $T_K = 150$\,K, as constrained by the rotational diagram.

In the model, the line-width will increase from $\sim 4$\,kms$^{-1}$ at the edge of the structure emission at $4\farcs$ to $\sim 9$\,kms$^{-1}$ at the center of the cavity. This will be implemented with a linear velocity gradient.

The fractional abundance of SiO is one of the most uncertain parameters. We will use the value estimated in Section~\ref{sec:abundances}, $X_{\rm SiO} \sim 10^{-8}$. Since the SiO emission comes from the outflow and cavity regions, we cannot constrain the fractional abundance of the cold ambient core gas with the same observations. A value of $10^{-11}$ is used, considering that a value of $\sim 10^{-12}$ is cited as the abundance of SiO in cold dense/starless cores environments \citep{schilke_1997}.

\subsection{Radiative transfer model} \label{sec:radiative_transfer_model} To simulate the response of the interferometer, the resulting simulated data cubes were processed through the Common Astronomy Software Applications (CASA, see Section~\ref{sec:observations}) tasks \emph{simobserve}, to simulate a set of measured visibilities with the compact configuration of the ALMA array; and \emph{simanalyze}, to produce synthetic deconvolved images from the visibilities.

The high-velocity wings in the SiO spectra makes evident the presence of an outflow with $\sim 70$\,kms$^{-1}$ from the systemic velocity of the cloud. The model that we will use for the outflow is a cone for each lobe. The estimation of the dimensions of this cone is performed in the following way: the spatial offset between the peaks of the $\pm 40$\,kms$^{-1}$ spectral channels in the SiO emission is $\sim 1\farcs2$. The model from Section~\ref{sec:ring_model} indicates that the cavity has an inclination angle with respect to the line of sight of $\sim 8-10^{\circ}$. Assuming that the cavity and the outflow axis are aligned, the total extension of both outflow lobes is $\sim 0.29$\,pc at the source distance along its axis. Therefore, the height of the cone is 0.145\,pc. To estimate the opening angle (or equivalently, the base of the cone), we use the fact that the velocity at which the outflow starts in the red wing is $-79.9$\,kms$^{-1}$. At this velocity, the ring of emission is $\sim 1\farcs1$ in radius. The base of the outflow cone has a diameter of $2\farcs2$. This gives an opening angle of $\sim 30^{\circ}$.

On the inside of an outflow, the density is low and the velocity and temperature are high. On the contrary, the core/envelope gas has high density and low temperature and velocity. \citet{rawlings_2004} modeled an outflow with an inside-outflow density of $10^3$\,cm$^{-3}$. \citet{zhang_2013} simulated radiative transfer and SEDs of massive star formation and considered the effects of outflows. Their densities inside the outflow are $\sim 10^{2-3}$\,cm$^{-3}$. Close to the outflow axis, where the high-velocity collimated jet is located, the density can be somewhat higher. On the shocked region, close to the edge of the cone where the high-velocity flow is interacting with the quiescent core, we know that the pre-shock density is close to $10^6$\,cm$^{-3}$ for our outflow source, as calculated in Section~\ref{sec:sio_shock}. A shock has an enhancement of 10-100 times the pre-shock density \citep{draine_1993}. The size of a layer of very dense and shocked gas is taken as $\Delta_s = 5\times10^{15}$\,cm $\sim 0.002$\,pc \citep{gusdorf_2008}, the typical length of a shock. One constrain we have for the density is the total mass of the outflow. Each lobe has a total mass of $\sim 24$\,$M_{\odot}$ \citep{bronfman_2008}, from lower resolution CO observations. The total mass will be estimated here by adding concentric disks approximating the shape of the cone

\begin{equation}
	M_{T} = \mu m_{H} 2 \pi  \sum_{z_i} \Delta z(z_i) \int_0^{\psi_{cone}} n(\psi) \psi d\psi \text{,}
\end{equation}

where $\psi$ is the cylindrical radius, $\psi_{cone}$ is the exterior cylindrical radius of the cone and $\Delta z$ is the height of the disk at each particular z height.

We will set a dense outflow region close to the edge of the cone. This will be limited by $\psi_0 = \psi_{cone} - \Delta_s$ for each particular $z$. The density is 

\small
\begin{equation}
    n(\psi)/\text{cm}^{-3}= 
\begin{cases}
    10^{2.0 + \frac{3.5}{(\psi_0/2)^4} (\psi - \psi_0/2)^4} 		& \text{if } \psi < \psi_0 \\
    10^{5.5} + 10^{5.5} enh \frac{1}{1 + \exp(-1E10(\psi - \psi_0))}  & \text{if } \psi >= \psi_0 \text{,}
\end{cases}
\end{equation}

\begin{figure*}
  \plotone{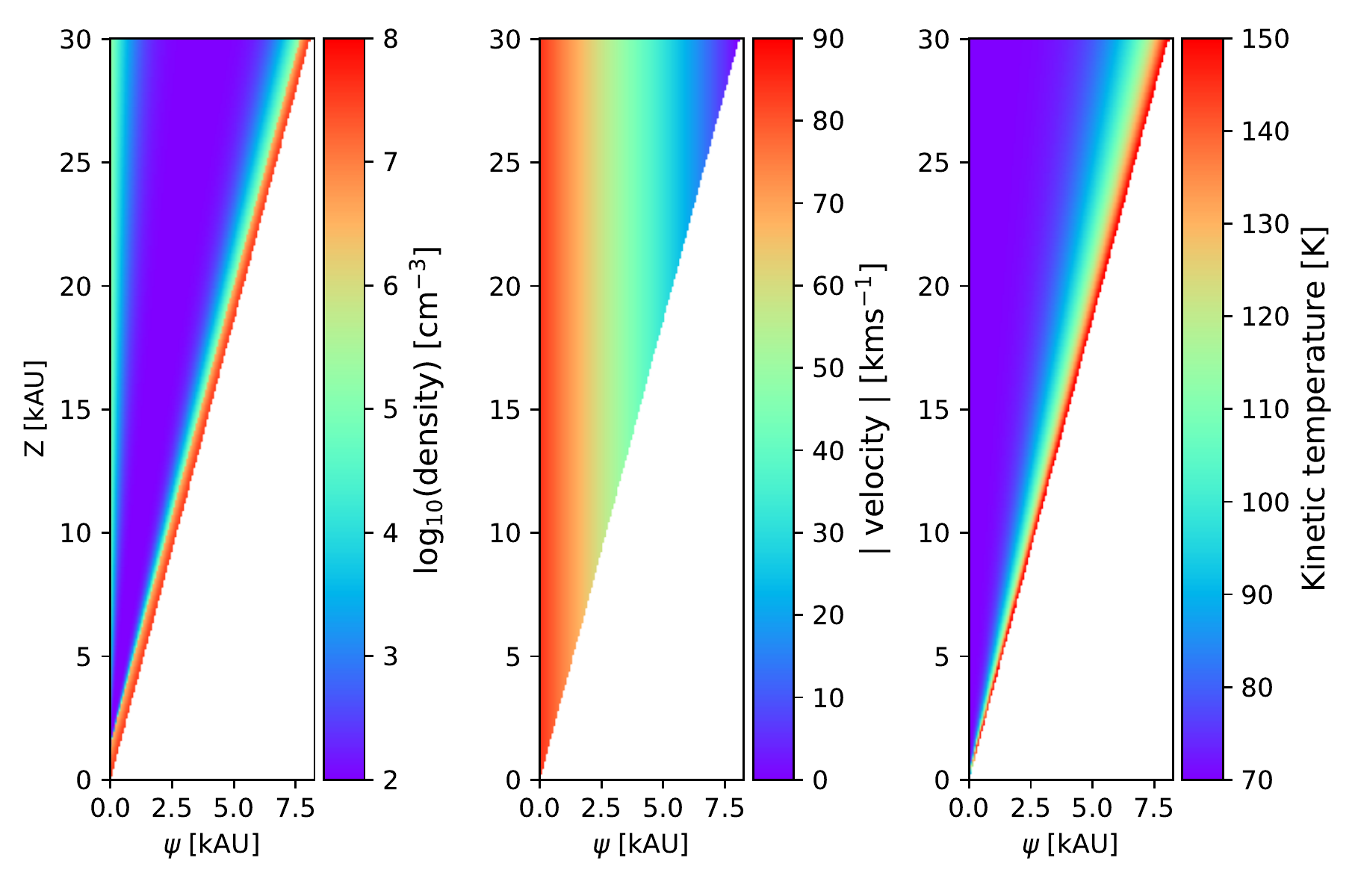}
  \caption{Left: Density distribution in cylindrical coordinates, used to simulate the outflow. The model is axysimmetric. Center: The same as the left, but showing the kinetic temperature profile. Right: The same as the left, but showing the module of the velocity field at each location. The velocity vector has a $\pm$Z direction.\label{fig:outflow_model_profiles}}
\end{figure*}

\normalsize
where $enh$ is the enhancement factor of the density, from 10 to 100 in a shock length of 0.00162\,pc. In this way, the density both close to the outflow axis and close to the cone limit is $\sim 10^{5.5}$\,cm$^{-3}$; meanwhile, in the middle of the outflow, it is nearly uniform and equal to $10^2$\,cm$^{-3}$. Finally, very close to the outflow limit there is a layer where the density is increased from 10 to 100 times the pre-shock density. Figure~\ref{fig:outflow_model_profiles} (Left) shows the density profile. The total mass of each lobe using this density is $\sim 21 M_{\odot}$. 

The kinematics of molecular outflows is discussed in \citet{stahler_1994}. In order to explain the PV plots of observed outflows, the so called ``outflow Hubble law'', Stahler proposes a series of equations characterizing the velocity distribution of an outflow. One conclusion is that the velocity field in an outflow depends on the cylindrical radius (if the Z-axis is the outflow symmetry axis). So the picture is that the highest velocities are found close to the outflow axis and then it decreases with the cylindrical radius. This picture is consistent with a high-velocity, highly collimated jet, close to the outflow axis. The velocity field will depend on the cylindrical radius

\begin{equation}
\| \vec{v} \| = 85 \left( 1 - \frac{\psi/1\farcs0}{4\farcs0\tan(\pi/12)}  \right) \text{[kms$^{-1}$]} \text{,}
\end{equation}

where the cylindrical radius $\psi$ is expressed in arcsec and $4\farcs0 \tan(\pi/12)$ corresponds to the radius of the base of the cone. The velocity vector will have a direction in the $\pm$Z-axis. The modulus of the velocity at each location is shown in Fig.~\ref{fig:outflow_model_profiles} (Center).

Using our estimates of temperature and the fact that outflows are somewhat hot compared with ambient core gas, we have a temperature of $\sim 150-200$\,K for the shocked dense gas, and 70\,K for the harboring core, estimated from CH$_3$CCH. For comparison, the outflow in \citet{rawlings_2004} has a temperature of 50\,K. The temperature distribution we use has $\sim 70$\,K for most of the interior of the outflow and then increases to 150\,K in the edge, where the gas is shocked, dense and hot. The temperature will be given by $T(\theta) = 70 {\rm [K]} \exp(162(\theta/rad)^4)$, where $\theta$ is in radians and corresponds to elevation angle (outflow opening angle). The distribution is shown in Fig.~\ref{fig:outflow_model_profiles} (Right).

A model for the shell emission is made independently. The expanding shell is modeled as two concentric ellipsoids. We know they are not spheres because the azimuthal PV plots are sinusoidal, and therefore there is an inclination induced asymmetry. The ellipsoids are oblate spheroids, with an aspect ratio of 1:1.5. The dimensions are $1\farcs8$ (inner ellipsoid) and $2\farcs4$ (outer ellipsoid) in the plane perpendicular to the outflow axis. 

The density of the ambient core, i.e. everything surrounding the cavity and expanding shell, depends on the radius, following eq.~\ref{eq:core_density}. The inner cavity is blown-up by the stellar winds, so its density will be similar to the insides of the outflow lobe. The density profile is given by \small 
\begin{equation}
    \frac{n(r)}{\text{cm}^{-3}}= 
\begin{cases}
    10^{2.0 + \frac{4.85}{(r_0/2)^4} (r - r_0/2)^4} 						& \text{if } r < r_0 \\
    10^{4.9} + 10^{4.85} enh \frac{1}{1 + \exp(-1E10(r - r_0))}   & \text{if } r >= r_0 \text{,}
\end{cases}
\end{equation} \normalsize where $r_0$ is the radius at which the most dense and shocked zone starts, located at the exterior limit of the inner cavity. It is defined by $r_0 = r_{\rm ellipsoid,inner} - \Delta_s$. The density in the expanding shell, i.e. between the inner and outer ellipsoids, is given by a power-law such that the border condition between the shell and the ambient core is fulfilled, that is, at $r = r_{\rm ellipsoid,outer}$, the density is given by $n_{\rm max}(r/r_{\rm ellipsoid,inner})^{-p_{\rm shell}} = 2.0 \times 10^6 [\text{cm}^{-3}] (r/1\farcs6)^{-0.8} $, where $n_{\rm max}$ is the maximum density reached at the cavity $r = r_{\rm ellipsoid,inner}$. The power-law index of the density of the shell is therefore given by

\small
\begin{equation}
	p_{\rm shell} = \frac{\log(2\times10^6) - 0.8\log(r_{\rm ellipsoid,outer}/1\farcs6) - \log(n_{\rm max}) }{-\log(r_{\rm ellipsoid,outer}/r_{\rm ellipsoid,inner})} \text{.}
\end{equation}

\begin{figure*}
	\plotone{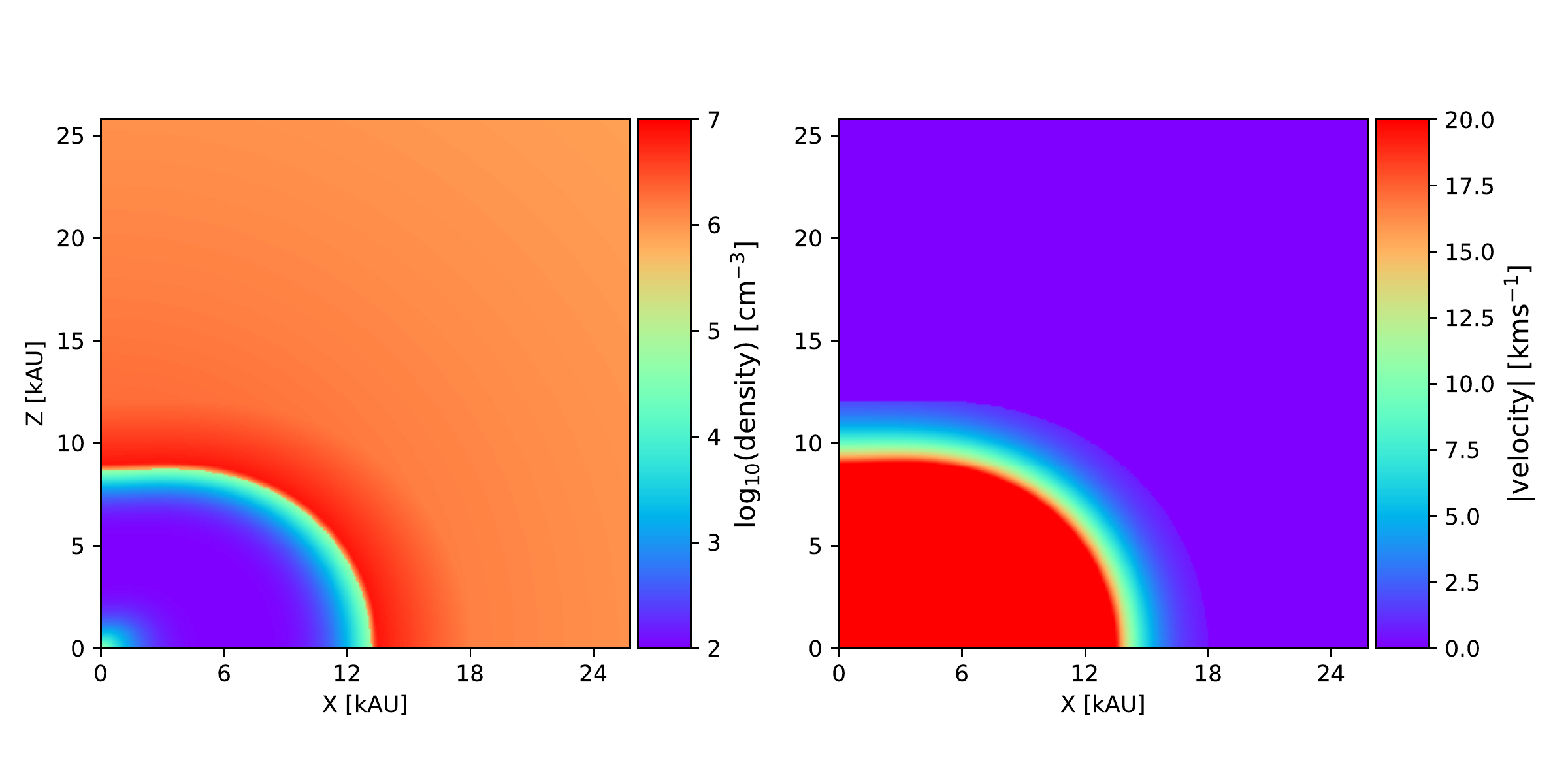}
	\caption{Left: Density profile of the modeled cavity, shell and ambient core. The Z axis is the same axis as in Fig. \ref{fig:outflow_model_profiles}, the symmetry axis of the outflow. Right: The same as the left, but showing the magnitude of the velocity vector field. The direction of the velocity vector is radially spherical.\label{fig:cavity_model_profiles}}
\end{figure*}

\normalsize
The full density profile is shown in Fig.~\ref{fig:cavity_model_profiles}. The mass of the cavity plus the expanding shell using this density description is $\sim 30 M_{\odot}$. 

Since the cavity is presumed to be blown-up by the stellar radiation output, the velocity will be set to the maximum value observed in the cavity, that is $20$\,kms$^{-1}$ (see Section~\ref{sec:ring_model}). The velocity is expected to diminish from the maximum value at $r=r_{\rm ellipsoid,inner}$ to the systemic value, i.e. 0\,kms$^{-1}$, within the expanding shell, that is at $r=r_{\rm ellipsoid,outer}$. This is because the ambient core is expected to be at the systemic velocity. To accomplish this, the modulus of the radial velocity is an exponential law, given by 

\begin{equation}
	| v | = v_{\rm max} \exp( - A_0 (r - r_{\rm ellipsoid,inner})) \text{,}
\end{equation} 

where $A_0$ is a constant that scales how fast the magnitude drops to zero. With this law, the velocity will be maximum at the edge of the cavity (or the inner ellipsoid of the expanding shell) and will approach zero (or the systemic velocity) at the outer edge of the expanding shell. The adopted value for $A_0$ is 160. The modulus of the radial velocity is shown in Fig.~\ref{fig:cavity_model_profiles} (Right).

The temperature inside the cavity is set to 70\,K. The enhanced-density layer close to the edge has a temperature of 150\,K. The expanding cavity and outer core have a temperature of 70\,K.

\end{document}